\documentclass[10pt,twocolumn,letterpaper]{article}

\usepackage{iccv}
\usepackage{times}
\usepackage{epsfig}
\usepackage{graphicx}
\usepackage{amsmath}
\usepackage{amssymb}

\usepackage{subcaption}
\usepackage{mwe}
\usepackage{booktabs}

\usepackage{soul}
\usepackage{xcolor}
\usepackage{courier}

\DeclareMathOperator*{\argmin}{arg\,min}

\newcommand{\matr}[1]{\mathbf{#1}}

\renewcommand{\paragraph}[1]{\vspace{2mm} \noindent \textbf{#1}}

\usepackage[pagebackref=true,breaklinks=true,letterpaper=true,colorlinks,bookmarks=false]{hyperref}

 \iccvfinalcopy 

\def\iccvPaperID{2437} 

\ificcvfinal\fi
\begin{document}

\title{Unsupervised Deep Learning for Structured Shape Matching}

\author{Jean-Michel Roufosse\\
LIX, \'Ecole Polytechnique\\
{\tt\small jm.roufosse@gmail.com}
\and
Abhishek Sharma\\
LIX, \'Ecole Polytechnique\\
{\tt\small kein.iitian@gmail.com}
\and
Maks Ovsjanikov\\
LIX, \'Ecole Polytechnique\\
{\tt\small maks@lix.polytechnique.fr}
}

\maketitle

\begin{abstract}
 We present a novel method for computing correspondences across 3D shapes using 
  unsupervised learning. Our method computes a non-linear transformation of given descriptor functions, while optimizing for global structural
  properties of the resulting maps, such as their bijectivity or approximate isometry. To this end, we use the functional maps framework, and build upon the recent FMNet architecture for descriptor learning. Unlike that approach, however, we show that learning can be done in a purely \emph{unsupervised setting}, without having access to any ground truth correspondences. This results in a very general shape matching method that we call SURFMNet for Spectral Unsupervised FMNet, and which can be used to establish correspondences within 3D shape collections without any prior information. We demonstrate on a wide range of challenging benchmarks, that our approach leads to state-of-the-art results compared to the existing unsupervised methods and achieves results that are comparable even to the supervised learning techniques. Moreover, our framework is an order of magnitude faster, and does not rely on geodesic distance computation or expensive post-processing.
\end{abstract}

\section{Introduction}

\begin{figure}[t!]
        \centering
        \begin{subfigure}[b]{0.43\columnwidth}
            \centering
            \includegraphics[width=0.95\textwidth, scale =.5]{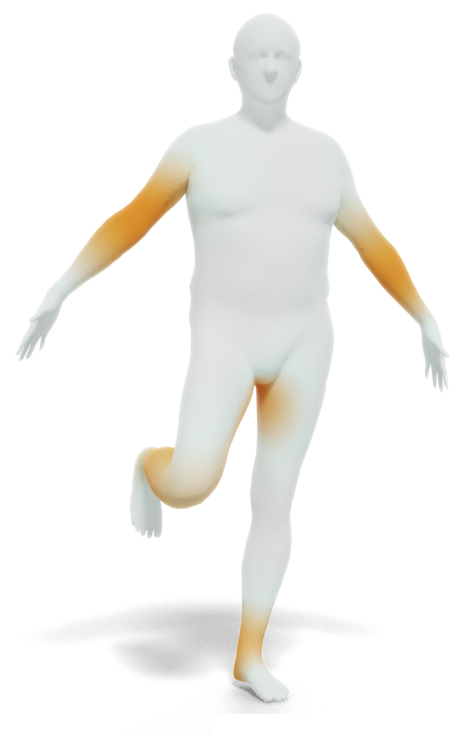}
            \caption*%
            {{Source descriptor before}}    
            \label{fig:FMnet.color}
        \end{subfigure}
        \hfill
        \begin{subfigure}[b]{0.43\columnwidth}  
            \centering 
            \includegraphics[width=0.95\textwidth,scale =.5]{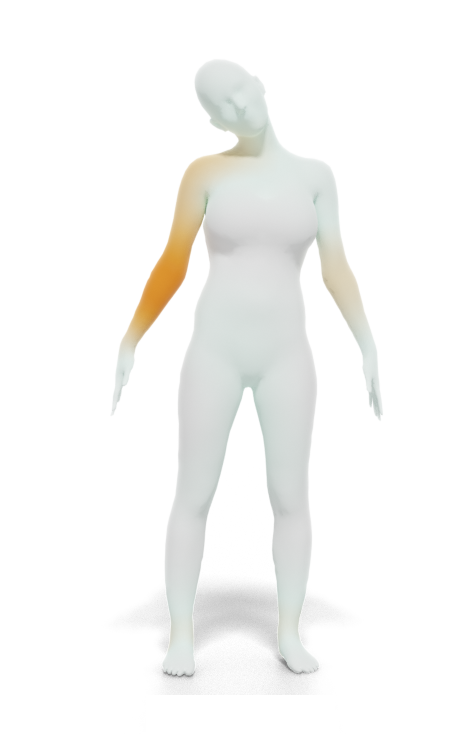}
            \caption*%
            {{ Target descriptor before }}    
            \label{fig:JustE3.color}
        \end{subfigure}
        \vskip\baselineskip
        \begin{subfigure}[b]{0.43\columnwidth}   
            \centering 
            \includegraphics[width=0.95\textwidth,scale =.5]{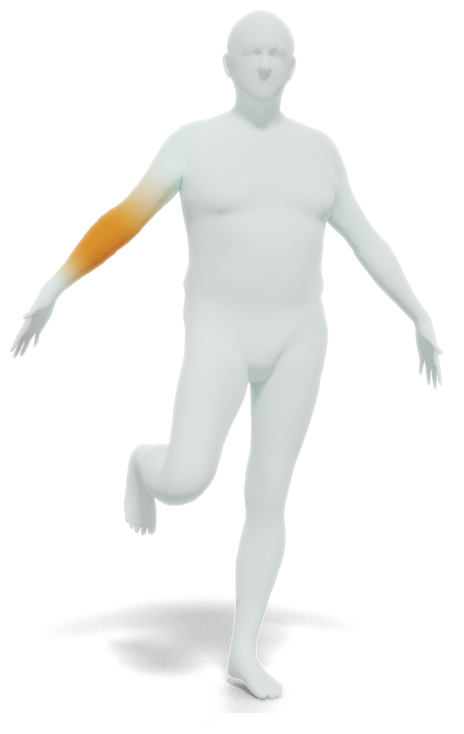}
            \caption*%
            {{ Source descriptor after}}   
            \label{fig:OursAll.color}
        \end{subfigure}
        \quad
        \begin{subfigure}[b]{0.43\columnwidth}   
            \centering 
            \includegraphics[width=0.95\textwidth,scale =.5]{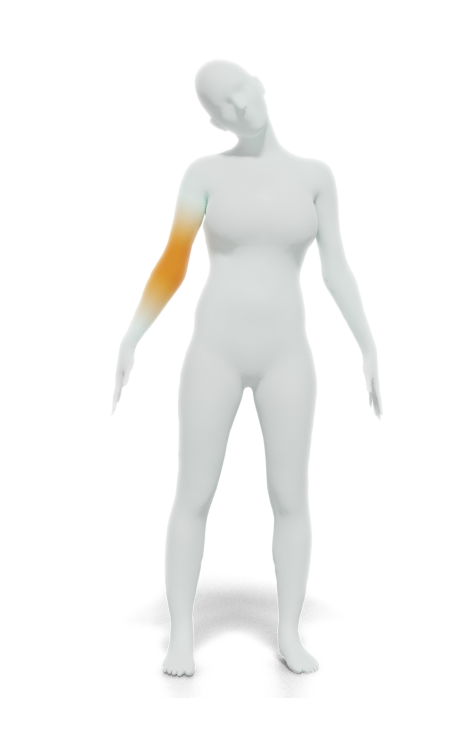}
            \caption*%
               {{ Target descriptor after}}   
            \label{fig:OursSubset.color}   
        \end{subfigure}
        
        \caption[]
        {Given a pair of shapes with noisy descriptors (top), our approach makes them more consistent (bottom) without the knowledge of the underlying map, and automatically computes an accurate pointwise correspondence.} 
        \label{fig:ComparingColor}
        \vspace{-3mm}
    \end{figure}

Shape matching is a fundamental problem in computer vision and geometric data
analysis, with applications in deformation transfer
\cite{sumner2004deformation} and statistical shape modeling \cite{bogo2014} among other domains. During the past decades, a large number of techniques have been proposed for both rigid and non-rigid shape matching \cite{tam2013registration}. The latter
case is both more general and more challenging since the shapes can potentially
undergo arbitrary deformations
(See Figure \ref{fig:ComparingColor}), which are not easy to characterize by purely
axiomatic approaches. As a result, several recent 
learning-based techniques have been proposed for addressing the shape correspondence problem,
e.g. \cite{corman2014supervised,litman2014learning,masci2015geodesic,wei2016dense}
among many others. Most of these approaches are based on the idea that the
underlying correspondence model can be learned from data, typically given in the
form of ground truth correspondences between some
shape pairs. In the simplest case, this can be formulated as a labeling problem, where different points,
e.g., in a template shape, correspond to labels to be predicted \cite{wei2016dense,monti2017}. More recently, several methods have been proposed for \emph{structured map
  prediction}, aiming to infer an entire map, rather than labeling each point
independently \cite{corman2014supervised,litany2017deep}. These techniques are
based on learning pointwise descriptors, but, crucially, impose a penalty on the
entire map, obtained using these descriptors, resulting in 
higher quality, globally consistent correspondences.
Nevertheless, while learning-based methods have achieved impressive performance, their utility is severely limited by requiring the presence of
high-quality ground truth maps between a sufficient number of training examples. This makes it difficult to apply such approaches to new shape
classes for which ground truth data is not available.

In our paper, we show that this limitation can be lifted and propose a purely
unsupervised strategy, which combines the accuracy of
learning-based methods with the generality of axiomatic techniques for
 shape correspondence. The key to our approach is a bi-level optimization
scheme, which optimizes for descriptors on the shapes, but imposes a penalty on the entire
map, inferred from them. For this, we use the recently proposed FMNet
architecture \cite{litany2017deep}, which exploits the functional map
representation \cite{ovsjanikov2012functional}. However, rather than penalizing
the deviation of the map from the ground truth, we enforce structural properties
on the map, such as its bijectivity or approximate isometry. This results in a
 shape matching method that achieves state-of-the-art accuracy among unsupervised methods and, perhaps surprisingly, achieves comparable performance even to supervised techniques.
\section{Related Work}
Computing correspondences between 3D shapes is a very well-studied area of computer vision and computer graphics. Below we only review the most closely related methods and refer the interested readers to recent surveys including \cite{van2011survey,tam2013registration,biasotti2016recent} for more in-depth discussions.

\paragraph{Functional Maps} 
Our method is built on the functional map representation, which was originally introduced in \cite{ovsjanikov2012functional} for solving non-rigid shape matching problems, and then extended significantly in follow-up works, including \cite{aflalo2013spectral,kovnatsky2013coupled,kovnatsky2015functional,burghard2017embedding,ezuz2017deblurring,rodola2017partial} among many others (see also \cite{ovsjanikov2017course} for a recent overview). 

One of the key benefits of this framework is that it allows us to represent maps between shapes as small matrices, which encode relations between basis functions defined on the shapes. Moreover, as observed by several works in this domain \cite{ovsjanikov2012functional,rustamov13,kovnatsky2013coupled,rodola2017partial, burghard2017embedding}, many natural properties on the underlying pointwise correspondences can be expressed as objectives on functional maps. This includes orthonormality of functional maps, which corresponds to the local area-preservation nature of pointwise correspondences \cite{ovsjanikov2012functional,kovnatsky2013coupled,rustamov13}; commutativity with the Laplacian operators, which corresponds to intrinsic isometries \cite{ovsjanikov2012functional}, preservation of inner products of gradients of functions, which corresponds to conformal maps \cite{rustamov13,burghard2017embedding,wang2018vector}; preservation of \emph{pointwise products} of functions, which corresponds to functional maps arising from point-to-point correspondences \cite{nogneng2017informative,nogneng2018improved}; and slanted diagonal structure of functional map in the context of partial shapes \cite{rodola2017partial,litany2017fully} among others.

Similarly, several other regularizers have been proposed, including exploiting the relation between functional maps in different directions \cite{eynard2016coupled}, the map adjoint \cite{huang2017adjoint}, and powerful cycle-consistency constraints \cite{huang2014functional} in shape collections to name a few. More recently constraints on functional maps have been introduced to promote map \emph{continuity} \cite{ren2018continuous,poulenard2018topological} and kernel-based techniques for extracting more information from given descriptors  \cite{wang2018kernel} among others.
All these methods, however, are based on combining first-order penalties that arise from enforcing \emph{descriptor preservation constraints} with these additional desirable structural properties of functional maps. As a result, any artefact or inconsistency in the pre-computed descriptors will inevitably lead to severe map estimation errors. Several methods have been suggested to use robust norms \cite{kovnatsky2013coupled,kovnatsky2015functional}, which can help reduce the influence of certain descriptors but still does not control the global map consistency properties.

Most recently, a powerful technique BCICP, for map optimization, was introduced in  \cite{ren2018continuous} that combines a large number of functional constraints with sophisticated post-processing, and careful descriptor selection. As we show below our method is simpler, more efficient and achieves superior accuracy even to this recent approach.

\paragraph{Learning-based Methods}
To overcome the inherent difficulty of axiomatic techniques, several methods have been introduced to learn the correct deformation model from data with learning-based methods. Some early approaches in this direction were used to learn either optimal parameters of spectral descriptors \cite{litman2014learning} or exploited random forests \cite{rodola2014dense} or metric learning \cite{cosmo2016matching} for learning optimal constraints given some ground truth matches.

More recently, with the advent of deep learning methods, several approaches have been proposed to learn transformations in the context of non-rigid shape matching. Most of the proposed methods either use Convolutional Neural Networks (CNNs) on depth maps, e.g. for dense human body correspondence \cite{wei2016dense} or exploit extensions of CNNs directly to curved surfaces, either using the link between convolution and multiplication in the spectral domain \cite{boscaini2015learning,defferrard2016convolutional}, or directly defining local parametrizations, for example via the exponential map, which allows convolution in the tangent plane of a point, e.g. \cite{masci2015geodesic,MasBosBroVan16,monti2017,poulenard2018multi} among others.

These methods have been applied to non-rigid shape matching, in most cases modeling it as a label prediction problem, with points corresponding to different labels. Although successful in the presence of sufficient training data, such approaches typically do not impose global consistency, and can lead to  artefacts, such as outliers, requiring post-processing to achieve high-quality maps.

\paragraph{Learning for Structured Prediction} 
 Most closely related to our approach are recent works that apply learning for structured map prediction \cite{corman2014supervised,litany2017deep}. These methods learn a transformation of given input descriptors, while optimizing for the deviation of the map computed from them using the functional map framework, from ground truth correspondences. By imposing a penalty on entire maps, and thus evaluating the ultimate use of the descriptors, these methods have led to significant accuracy improvements in practice. We note that concurrent to our work, Halimi et al. \cite{halimi2018self} also proposed an unsupervised deep learning method that computes correspondences without using the ground truth. This approach is similar to ours, but is based on computation of geodesic distances, while our method operates purely in the spectral domain making it extremely efficient.

\paragraph{Contribution}
Unlike these existing methods, we propose an \emph{unsupervised} learning-based approach that transforms given input descriptors, while optimizing for structural map properties, without any knowledge of the ground truth or geodesic distances. Our method, which can be seen as a bi-level optimization strategy, allows to explicitly control the interaction between pointwise descriptors and global map consistency, computed via the functional map framework. As a result, our technique is  scalable with respect to shape complexity, leads to significant improvement compared to the standard 
unsupervised methods, and achieves comparable performance even to supervised approaches.
\section{Background \& Motivation}
\subsection{Shape Matching and Functional Maps}
\label{subsec:fmaps}
Our work is based on the functional map framework and representation. For completeness, we briefly review the basic
notions and pipeline for estimating functional maps, and refer the interested reader to a recent course
\cite{ovsjanikov2017course} for a more in-depth discussion.


\paragraph{Basic Pipeline} Given a pair of shapes, $S_1,S_2$ represented as triangle meshes, and
containing, respectively, $n_1$ and $n_2$ vertices, the basic pipeline for computing a map between
them using the functional map framework, consists of the following main steps (see Chapter 2 in
\cite{ovsjanikov2017course}) :
\begin{enumerate}
  \setlength\itemsep{-0.2em}
\item Compute a small set of $k_1, k_2$ of basis functions on each shape, e.g. by taking the first few
  eigenfunctions of the respective Laplace-Beltrami operators.
\item Compute a set of descriptor \emph{functions} on each shape that are expected to be approximately preserved by the
  unknown map. For example, a descriptor function can correspond to a particular dimension (e.g. choice of time
  parameter of the Heat Kernel Signature \cite{hks}) computed at every point. Store their coefficients in the
  respective bases as columns of matrices $\matr{A}_1, \matr{A}_2$.
\item Compute the optimal \emph{functional map} $\matr{C}$ by solving the following optimization problem:
\begin{align}
\label{eq:opt_problem}
C_{\text{opt}} = \argmin_{\matr{C}_{12}} E_{\text{desc}}\big(\matr{C}_{12}\big) + \alpha E_{\text{reg}}\big(\matr{C}_{12}\big),
\end{align}
where the first term aims at the descriptor preservation: $E_{\text{desc}}\big(\matr{C}_{12}\big)
= \big\Vert \matr{C}_{12} \matr{A}_1 - \matr{A}_2\big\Vert^2$, whereas the second term regularizes
the map by promoting the correctness of its overall structural properties. The simplest approach penalizes the failure of the unknown functional map to commute with the Laplace-Beltrami operators:
        \begin{align}
        \label{eq:energy:laplacian_comm}
        E_{\text{reg}}(C_{12}) = \big\Vert \matr{C}_{12}\matr{\Lambda}_1 - \matr{\Lambda}_2 \matr{C}_{12} \big\Vert^2    
        \end{align}
         where $\matr{\Lambda}_1$ and $\matr{\Lambda}_2$ are diagonal matrices of the Laplace-Beltrami eigenvalues on the two shapes.
       \item Convert the functional map $\matr{C}$ to a point-to-point map, for example using nearest neighbor search in
         the spectral embedding, or using other more advanced techniques \cite{rodola-vmv15,ezuz2017deblurring}.
\end{enumerate}

One of the strengths of this pipeline is that typically Eq.~(\ref{eq:opt_problem}) leads to a simple (e.g., least
squares) problem with  $k_1 k_2$ unknowns, independent of the number of points on the shapes. This formulation has been
extended using e.g.  manifold optimization \cite{kovnatsky2016madmm}, descriptor preservation constraints via
commutativity \cite{nogneng2017informative} and, more recently, with kernelization \cite{wang2018kernel} among many others (see also
Chapter 3 in \cite{ovsjanikov2017course}).

\subsection{Deep Functional Maps}
Despite its simplicity and efficiency, the functional map estimation pipeline described above is fundamentally dependent
on the initial choice of descriptor functions. To alleviate this dependence, several approaches have been proposed to
learn the optimal descriptors from data \cite{corman2014supervised,litany2017deep}. In our work, we build upon a recent
deep learning-based framework, called FMNet, introduced by Litany et al. \cite{litany2017deep} that aims to transform a
given set of descriptors so that the optimal map computed using them is as close as possible to some ground truth map
given during training.

Specifically, the approach proposed in \cite{litany2017deep} assumes, as input, a set of shape pairs for which ground truth point-wise maps are known, and aims to solve the following problem:
\begin{align}
\label{eq:fmnet1}
&\min_{T} \sum_{(S_1,S_2) \in \text{Train}} l_F (Soft(\mathbf{C}_{\text{opt}}), GT_{(S_1, S_2)}), \text{ where }\\
\label{eq:fmnet2}
&\matr{C}_{\text{opt}} = \argmin_{\matr{C}} \| \matr{C} \matr{A}_{T(D_1)} - \matr{A}_{T(D_2)} \|.
\end{align}
Here $T$ is a non-linear transformation, in the form of a neural network, to be applied to some input descriptor functions $D$, $\text{Train}$ is the set of training pairs for which ground truth correspondence $GT_{(S_1, S_2)}$ is known, $l_F$ is the \emph{soft error loss}, which penalizes the deviation of the computed functional map $\mathbf{C}_{\text{opt}}$, after converting it to a soft map $Soft(\mathbf{C}_{\text{opt}})$ from the ground truth correspondence, and $\mathbf{A}_{T(D_1)}$ denotes the transformed descriptors $D_1$ written in the basis of shape~$1$. In other words, the FMNet framework \cite{litany2017deep} aims to learn a transformation $T$ of descriptors, so that the transformed descriptors $T(D_1)$, $T(D_2)$, \emph{when used within the functional map pipeline} result in a soft map that is as close as possible to some known ground truth correspondence. Unlike methods based on formulating shape matching as a labeling problem this approach evaluates the quality of the \emph{entire map}, obtained using the transformed descriptors, which as shown in \cite{litany2017deep} leads to significant improvement compared to several strong baselines.

\paragraph{Motivation}
Similarly to other supervised learning methods, although FMNet \cite{litany2017deep} can result in highly accurate correspondences, its applicability is limited to shape classes for which high-quality ground truth maps are available. Moreover, perhaps less crucially, the soft map loss in FMNet is based on the knowledge of geodesic distances between all pairs of points, making it computationally expensive. Our goal, therefore, is to show that a similar approach can be used more widely, without any training data, while working purely in the spectral domain.

\section{SURFMNet}
\subsection{Overview}
\label{subsec:overview}

In this paper, we introduce a novel approach, which we call \textbf{SURFMNet} for Spectral Unsupervised FMNet. Our method aims to optimize for non-linear
transformations of descriptors, in order to obtain high-quality functional, and thus pointwise
maps. For this, we follow the general strategy proposed in FMNet
\cite{litany2017deep}. 
\begin{figure}[t!]
\begin{center}
\includegraphics[scale=0.35]{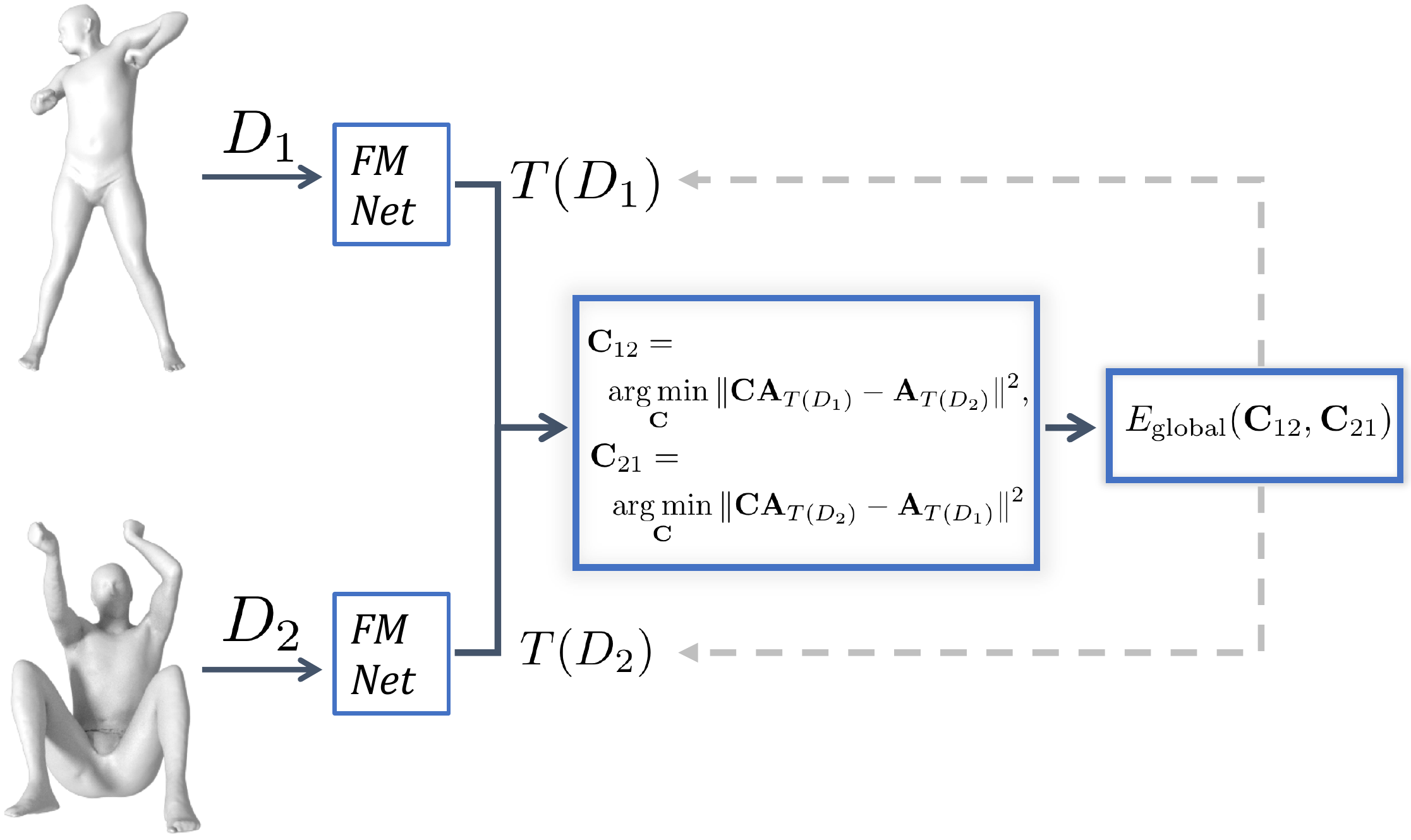}
\caption{Overview of our SURFMNet approach: given a pair of shapes and their descriptors $D_1$, $D_2$, we optimize for a non-linear transformation $T$ using the FMNet architecture so that the transformed descriptors lead to functional maps that best satisfy the structural constraints. 
\label{fig:network}
\vspace{-4mm}}
\end{center}
\end{figure}
However, crucially, rather than penalizing the deviation of the computed map
from the known ground truth correspondence, we evaluate the structural properties of the inferred
functional maps, such as their bijectivity or orthogonality. Importantly, we express all these
desired properties, and thus the penalties during optimization, purely in the spectral domain, which
allows us to avoid the conversion of functional maps to soft maps during optimization as was done in
\cite{litany2017deep}. Thus, in addition to being purely unsupervised, our approach is also more
efficient since it does not require pre-computation of geodesic distance matrices or expensive
manipulation of large soft map matrices during training. 

To achieve these goals, we build on the FMNet model, described in Eq.~(\ref{eq:fmnet1}) and (\ref{eq:fmnet2}) in several ways: first, we propose to consider functional maps in both directions,
i.e. by treating the two shapes as both source and target; second, we remove the conversion
from functional to soft maps; and, most importantly, third, we replace the soft map loss with
respect to ground truth with a set of penalties on the computed functional maps, which are described
in detail below. Our optimization problem can be written as:
\begin{align}
\label{eq:ours1}
&\min_{T} \sum_{(S_1,S_2)} \sum_{i \in \text{penalties}} w_i E_i(\mathbf{C}_{12}, \mathbf{C}_{21}), \text{ where }\\
\label{eq:ours2}
&\matr{C}_{12} = \argmin_{\matr{C}} \| \matr{C} \matr{A}_{T(D_1)} - \matr{A}_{T(D_2)} \|, \\ 
\label{eq:ours3}
&\matr{C}_{21} = \argmin_{\matr{C}} \| \matr{C} \matr{A}_{T(D_2)} - \matr{A}_{T(D_1)} \|.
\end{align}
Here, similarly to Eq.~(\ref{eq:fmnet1}) above, $T$ denotes a non-linear transformation in the form
of a neural network, $(S_1, S_2)$ is a set of pairs of shapes in a given collection, $w_i$ are
scalar weights, and $E_i$ are the penalties, described below. Thus, we aim to optimize for a
non-linear transformation of input descriptor functions, such that functional maps computed from
transformed descriptors possess certain desirable structural properties and are expressed via penalty
minimization. Figure~\ref{fig:network} illustrates our proposed method where 
we denote the total sum of all penalty terms in Eq. \eqref{eq:ours1} as $E_{\text{global}}$ and back-propagation via grey dashed lines. 

When deriving the penalties used in our approach, we exploit the links between properties of
functional maps and associated pointwise maps, that have been established in several previous works
\cite{ovsjanikov2012functional,rustamov13,eynard2016coupled,nogneng2017informative}. Unlike all
these methods, however, we \emph{decouple the descriptor preservation constraints from structural
  map properties}. This allows us to optimize for descriptor functions, and thus, gain a very strong
resilience in the presence of noisy or uninformative descriptors, while still exploiting the
compactness and efficiency of the functional map representation.



\subsection{Deep Functional Map Regularization}
In our work, we propose to use four regularization terms, by including them as a penalties in the objective function, all inspired by desirable map properties.

\paragraph{Bijectivity} Given a pair of shapes and the functional maps in both directions, perhaps the simplest
requirement is for them to be inverses of each other, which can be enforced by penalizing the difference between their composition and the identity map. This penalty, used for functional map estimation in \cite{eynard2016coupled}, can be written, simply  as:
%
\begin{align}
E_1 = \| \matr{C}_{12} \matr{C}_{21} - \matr{I}\|^2 + \| \matr{C}_{21} \matr{C}_{12} - \matr{I}\|^2
\end{align}


\paragraph{Orthogonality} As observed in several works \cite{ovsjanikov2012functional,rustamov13} a point-to-point map is locally area preserving if and only if the corresponding functional map is \emph{orthonormal}. Thus, for shape pairs,
approximately satisfying this assumption, a natural penalty in our unsupervised pipeline is:
%
\begin{align}
E_2 = \| \matr{C}_{12}^\top \matr{C}_{12} - \matr{I}\|^2 + \| \matr{C}_{21}^\top \matr{C}_{21} - \matr{I}\|^2
\end{align}

\paragraph{Laplacian commutativity} Similarly, it is well-known that a pointwise map is an intrinsic isometry if and
only if the associated functional map commutes with the Laplace-Beltrami operator
\cite{rosenberg1997laplacian,ovsjanikov2012functional}. This has motivated using the lack of commutativity as a
regularizer for functional map computations, as mentioned in Eq.~(\ref{eq:energy:laplacian_comm}). In our work, we use
it to introduce the following penalty:
%
\begin{align}
E_3 = \big\Vert \matr{C}_{12}\matr{\Lambda}_1 - \matr{\Lambda}_2 \matr{C}_{12} \big\Vert^2 + \big\Vert \matr{C}_{21}\matr{\Lambda}_2 - \matr{\Lambda}_1 \matr{C}_{21} \big\Vert^2
\label{eq:E3}
\end{align} where $\matr{\Lambda}_1$ and $\matr{\Lambda}_2$ are diagonal matrices of the Laplace-Beltrami eigenvalues on the two shapes.

\paragraph{Descriptor preservation via commutativity} 
The previous three penalties capture desirable properties of pointwise correspondences when
expressed as functional maps. Our last penalty promotes functional maps that arise from point-to-point maps, rather than more general soft correspondences. To achieve this, we follow the approach proposed in
\cite{nogneng2017informative} based on preservation of pointwise products of functions. Namely, it is known that a non-trivial linear transformation $\mathcal{T}$ across function spaces corresponds to a point-to-point map if and only if $\mathcal{T}(f \odot h) = \mathcal{T}(f) \odot \mathcal{T}(h) $ for any pair of functions $f,h$. Here $\odot$ denotes the pointwise product between functions \cite{Comp}, i.e. $(f \odot h) (x) = f(x) h(x)$. When $f$ is a descriptor function on the source and $g$ is the corresponding descriptor on the target, the authors of \cite{nogneng2017informative} demonstrate that this condition can be rewritten in the reduced basis as follows: 
$\matr{C} \matr{M}_{f} = \matr{M}_{g} \matr{C}$, where $\matr{M}_{f} = \Phi^{+}\text{Diag}(f)\Phi,$ and $\matr{M}_{g} = \Psi^{+}\text{Diag}(g)\Psi$. This leads to the following penalty, in our setting:
\begin{align}
\label{eq:descmult}
\begin{split}
 E_4 = \sum_{(f_i, g_i) \in \text{Descriptors}} || \matr{C}_{12} \matr{M}_{f_i} - \matr{M}_{g_i} \matr{C}_{12}||^2 \\ 
+ ||  \matr{C}_{21} \matr{M}_{g_i} - \matr{M}_{f_i} \matr{C}_{21} ||^2, \\
\matr{M}_{f_i} = \Phi^{+}\text{Diag}(f_i)\Phi, \matr{M}_{g_i} = \Psi^{+}\text{Diag}(g_i)\Psi .
\end{split}
\end{align}
In this expression, $f_i$ and $g_i$ are the \emph{optimized} descriptors on source and target shape, obtained by
the neural network, and expressed in the full (hat basis), whereas $\Phi, \Psi$ are the fixed basis
functions on the two shapes, and ${+}$ denotes the Moore-Penrose pseudoinverse.

\subsection{Optimization} As mentioned in Section~\ref{subsec:overview}, we incorporate these four penalties into the energy in Eq.~(\ref{eq:ours1}). Importantly, the only unknowns in this optimization are the parameters of
the neural network applied to the descriptor functions. The functional maps $\matr{C}_{12}$ and
$\matr{C}_{21}$ are fully determined by the optimized descriptors via the solution of the
optimization problems in Eq.~(\ref{eq:ours2}) and Eq.~(\ref{eq:ours3}).
Note that although stated as optimization problems, both Eq.~(\ref{eq:ours2}) and Eq.~(\ref{eq:ours3}) reduce to solving a linear system of equations. This is easily differentiable using the well-known closed-form expression for derivatives of matrix inverses \cite{petersen2012matrix}.
Moreover, the functionality of differentiating a linear system of equations is implemented in TensorFlow \cite{tensorflow2015-whitepaper} and we use it directly, in the same way as it was used in the original FMNet work. Finally, all of the penalties $E_1, E_2, E_3, E_4$ are differentiable with respect to the functional maps $\matr{C}_{12}, \matr{C}_{21}$. This means that the gradient of the total energy can be back-propagated to the neural network $T$ in Eq.~(\ref{eq:ours1}), allowing us to optimize for the descriptors while penalizing the structural properties of the functional maps.

\section{Implementation \& Parameters}
\paragraph{Implementation details} We implemented \footnote{Code available at \url{https://github.com/LIX-shape-analysis/SURFMNet}.} our method in TensorFlow \cite{tensorflow2015-whitepaper} by adapting the open-source implementation of
FMNet \cite{litany2017deep}. Thus, the neural network $T$ used for transforming descriptors in our approach, in
Eq.~(\ref{eq:ours1}) is exactly identical to that used in FMNet, as mentioned in Eq.~(\ref{eq:fmnet1}). Namely, this
network is based on a residual architecture, consisting of 7~fully connected residual layers with exponential linear
units, without dimensionality reduction. Please see Section~5 in \cite{litany2017deep} for more details.

\paragraph{}  Following the approach of FMNet \cite{litany2017deep}, we also sub-sample a random set of 1500 points at each training step, for efficiency. However, unlike their method, sub-sampling is done independently on each shape, without enforcing consistency. Remark that our network is fully connected on the dimensions of the descriptors, not across vertices themselves. For example, the first layer has $352 \times 352$ weights (not $1500 \times 352$ weights) where $352$ and 1500 are the dimensions of the SHOT descriptors, and no. of sampled vertices respectively. Indeed, in exactly the same way as in FMNet, our network is applied on the descriptors of each point independently, using the same (learned) weights, and  different points on the shape only communicate through the functional map estimation layer, and not in the MLP layers. This ensures invariance to permutation of shape vertices. We also randomly sub-sample~20\% of the optimized descriptors for our penalty~$E_4$ at each training step to
avoid manipulating a large set of operators. We observed that this sub-sampling not only
helps to gain speed but also robustness during optimization. Importantly, we do not form large diagonal matrices explicitly, but rather define the multiplicative operators $\matr{M}$ in objective $E_4$ directly via pointwise products and summation using contraction between tensors. 

Finally, we convert functional maps to pointwise ones with nearest neighbor search in the spectral domain, following the original approach \cite{ovsjanikov2012functional}.

\paragraph{Parameters}
Our method takes two types of inputs: the input descriptors, and the scalar weights $w_i$ in Eq. \eqref{eq:ours1}. In all
experiments below, we used the same SHOT \cite{shot} descriptors as in FMNet \cite{litany2017deep} with the same
parameters, which leads to a 352-dimensional vector per point, or equivalently, 352 ~descriptor functions on each shape. For the scalar weights, $w_i$, we used the same four fixed values for all experiments below (namely,~$w_1 = 10^3$,~$w_2 = 10^3$,~$w_3 = 1$ and~$w_4 = 10^5$), which were obtained by examining the relative penalty values obtained throughout the
optimization on a small set of shapes, and setting the weights inversely proportionally to those values. We train our network with a batch size of 10 for 10 000 iterations using a learning rate of 0.001 and ADAM optimizer \cite{adam}.

\section{Results}



\begin{table}
\begin{center}

\resizebox{.48\textwidth}{!}
{\begin{tabular}{l|ccccc}\toprule
\textbf{Methods} & E1+E2+E3+E4 & E3  & E1   & E2   & E4  \\ \midrule
\textbf{Geodesic Error}& 0.020  & 0.073  & 0.083     & 0.152 & 0.252 
\end{tabular}}
\end{center}
\vspace{-4mm}
 \caption{Ablation study of penalty terms in our method 
  on the FAUST benchmark.
\label{tab:ablation}}
\vspace{-5mm}
\end{table}

\paragraph{Datasets}
We evaluate our method on the following datasets: the original FAUST dataset \cite{bogo2014} containing 100 human
shapes in 1-1 correspondence and the remeshed versions of SCAPE \cite{scape} and FAUST \cite{bogo2014} datasets, made publicly available recently by Ren et al. \cite{ren2018continuous}. These datasets  were obtained by independently re-meshing each shape to approximately 5000 vertices using the LRVD re-meshing method \cite{yan2014low}, while keeping track of the ground truth maps within each collection. This results in meshes that
are no longer in 1-1 correspondence, and indeed can have different number of vertices. 
The re-meshed datasets therefore offer significantly more variability in terms of shape structures, including e.g. point sampling density, making them more challenging for existing algorithms. Let us note also that the SCAPE dataset is slightly more challenging since the shapes are less regular (e.g., there are often reconstruction artefacts on hands and feet) and have fewer features than those in FAUST.

We stress that although we also evaluated on the original FAUST dataset, we view the remeshed datasets as more
realistic, providing a more faithful representation of the accuracy and generalization power of different
techniques.

\begin{figure*}[t!]
\begin{center}
\begin{subfigure}{.25\paperwidth}
  \centering
  \includegraphics[trim={.5cm 0 1.5cm 0}, clip, width=\linewidth]{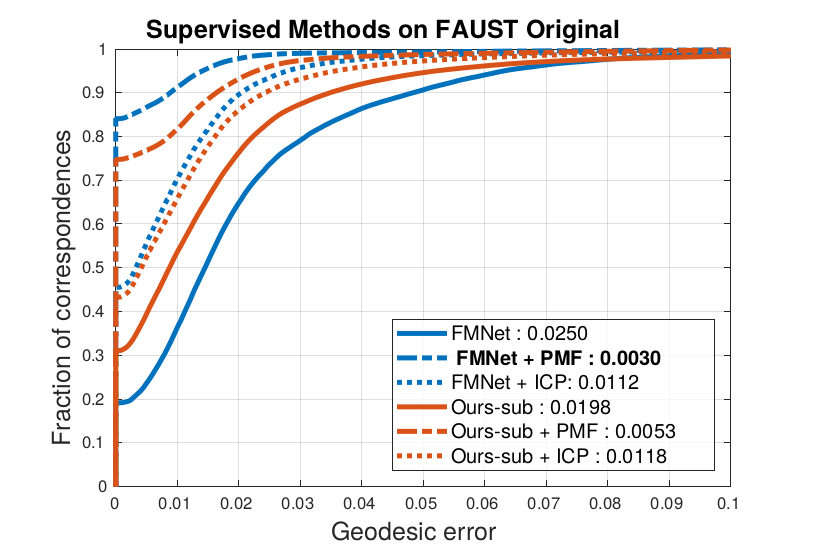}
  \label{fig:sfig1}
\end{subfigure}%
\begin{subfigure}{.25\paperwidth}
  \centering
  \includegraphics[trim={.5cm 0 1.5cm 0}, clip,width = \linewidth]{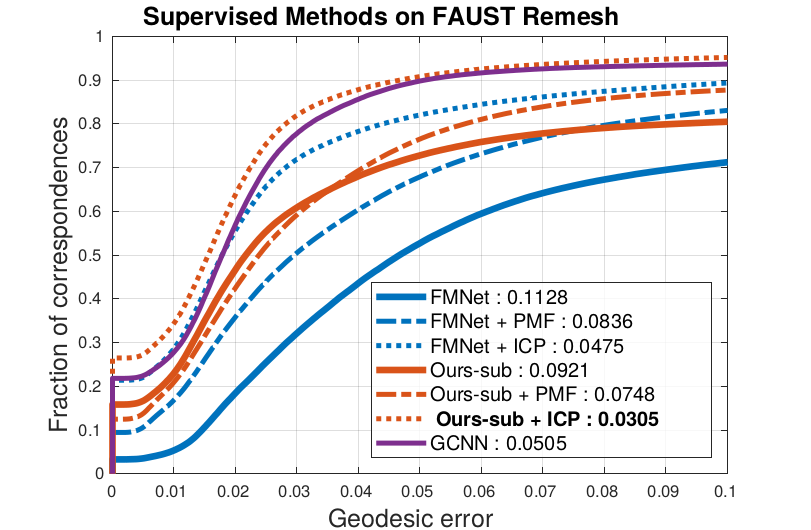}
  \label{fig:sfig2}
\end{subfigure}
\begin{subfigure}{.25\paperwidth}
  \centering
  \includegraphics[trim={.5cm 0 1.5cm 0}, clip, width = \linewidth]{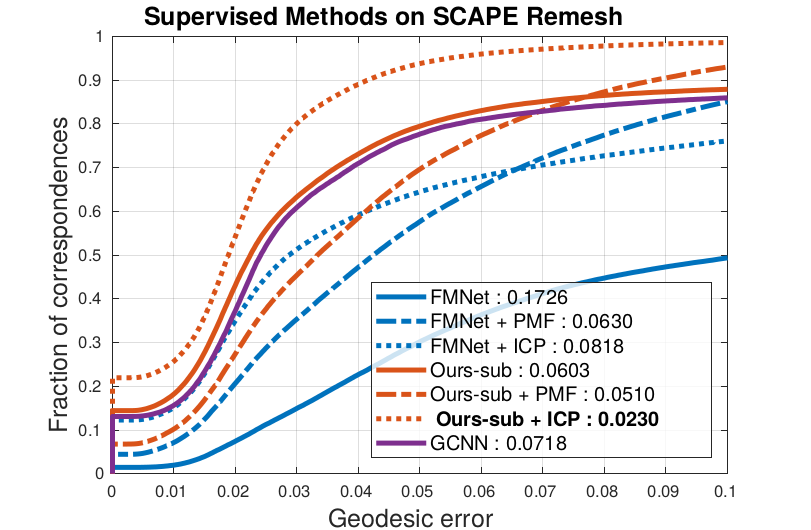}
  \label{fig:sfig2}
\end{subfigure}
\end{center}
\vspace{-8mm}
   \caption{Quantitative evaluation of pointwise correspondences comparing our method with Supervised Methods.
   \vspace{-2mm}}
\label{fig:sup-plot}
\end{figure*}

\begin{figure*}[t!]
\begin{center}
\begin{subfigure}{.25\paperwidth}
  \centering
  \includegraphics[trim={.5cm 0 1.5cm 0}, clip, width=\linewidth]{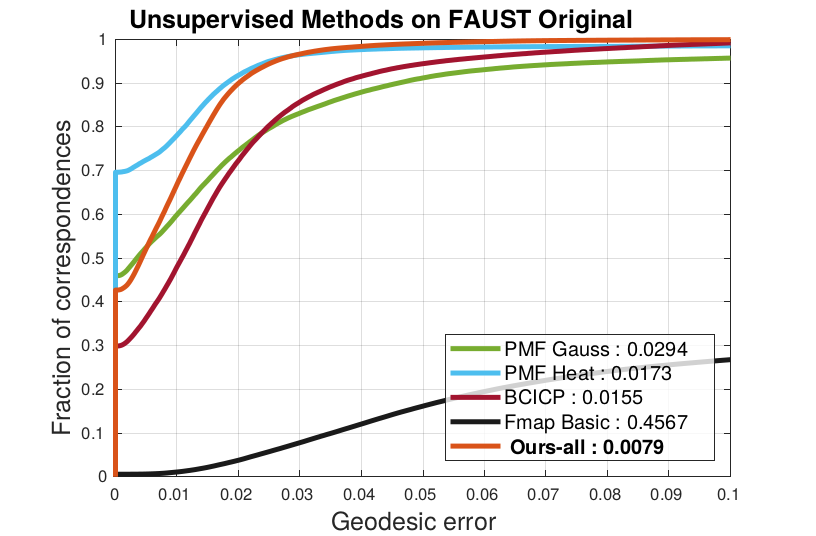}
  \label{fig:sfig1}
\end{subfigure}%
\begin{subfigure}{.25\paperwidth}
  \centering
  \includegraphics[trim={.5cm 0 1cm 0}, clip, width = \linewidth]{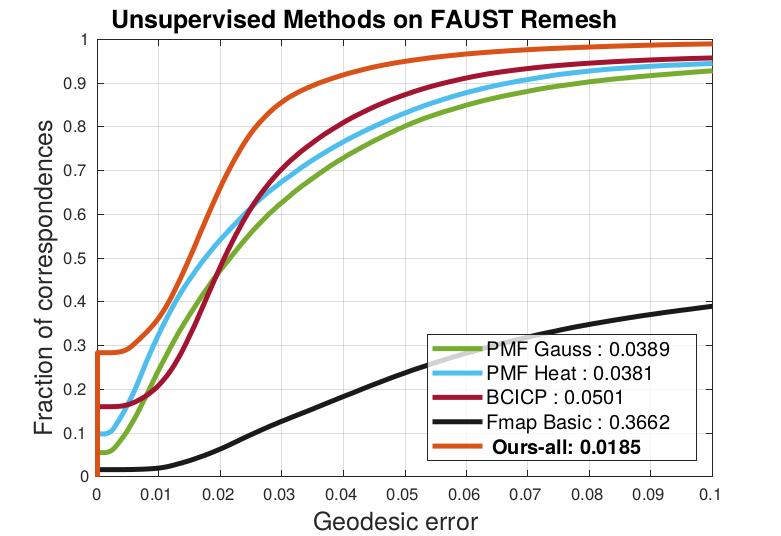}
  \label{fig:sfig2}
\end{subfigure}
\begin{subfigure}{.25\paperwidth}
  \centering
  \includegraphics[trim={.5cm 0 1.5cm 0}, clip, width = \linewidth]{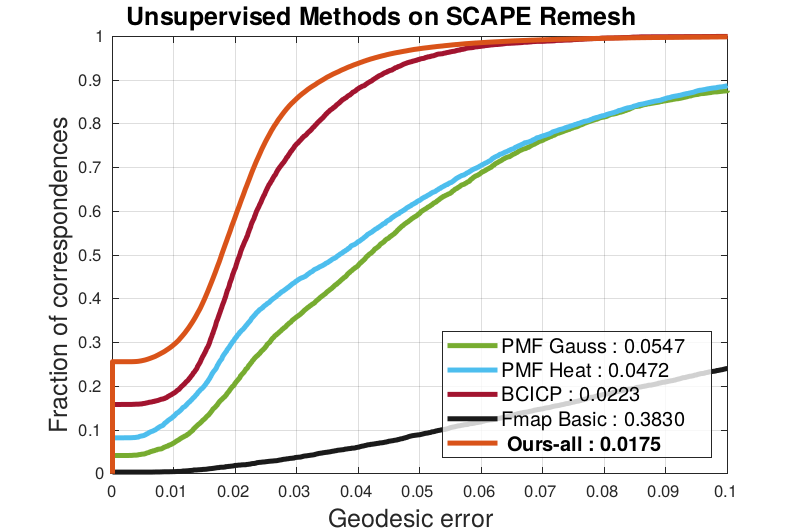}
  \label{fig:sfig2}
\end{subfigure}
\end{center}
\vspace{-8mm}
   \caption{Quantitative evaluation of pointwise correspondences comparing our method with Unsupervised Methods. 
   \vspace{-2mm}}
\label{fig:unsup-plot}
\end{figure*}

\paragraph{Ablation study} We first evaluated the relative importance of the different
penalties in our method on the FAUST shape dataset \cite{bogo2014}. We evaluated the average correspondence geodesic error  with respect to the ground truth maps. 

Table~\ref{tab:ablation} summarizes the quality of the computed correspondences between shapes in the test set, using different combination of penalties. 
We observe that the combination of all four penalties significantly out-performs any other subsets. Besides, among individual penalties used independently, the Laplacian commutativity gives the best result. For more combinations of penalty terms, we refer to a more detailed ablation study in the supplementary material.

\paragraph{Baselines}
We compared our method to several techniques, both supervised and fully automatic. For conciseness, we refer to \textbf{SURFMNet} as \textbf{Ours} in the following text. For a fair comparison with FMNet, we evaluate our method in two settings: \textbf{Ours-sub} and \textbf{Ours-all}. For Ours-sub, we split each dataset into training and test sets containing 80 and 20 shapes respectively, as done in \cite{litany2017deep}. 
For Ours-all, we optimize over all the dataset and apply the optimized network on the same test set as before. We stress that unlike FMNet, our method does not use any ground truth in either setting. We use the notation Ours-sub only to emphasize the split of dataset into train and test since the ``training set'' was only used for descriptor optimization with the functional map penalties introduced above without any ground truth. 

Since the original FMNet work \cite{litany2017deep} already showed very strong improvement compared to existing supervised learning methods we primarily compare to this approach. For reference, we also compare to the Geodesic Convolutional Neural Networks (GCNN) method of \cite{masci2015geodesic} on the remeshed datasets, which were not considered in \cite{litany2017deep}. GCNN is a representative supervised method based on local shape parameterization, and as FMNet assumes, as input, ground truth maps between a subset of the training shapes. For supervised methods, we always split the datasets into 80 (resp. 60) shapes for training and 20 (resp. 10) for testing in the FAUST and SCAPE datasets respectively. 

Among fully automatic methods, we use the Product Manifold Filter method with
the Gaussian kernel \cite{vestner2017product} (PMF Gauss) and its variant with the Heat kernel \cite{vestnerefficient} (PMF Heat). We also compare to the recently proposed BCICP \cite{ren2018continuous}, which achieved state-of-the-art results among axiomatic methods. With a slight abuse of notation, we denote these non-learning methods as Unsupervised in Figure~\ref{fig:unsup-plot} since none of these methods use ground truth. 
Finally, we also evaluated the basic functional map approach, based on directly optimizing the functional maps as outlined in Section~\ref{subsec:fmaps}, but using all four of our energies for regularization. This method, which we call ``Fmap Basic'' can be viewed as a combination of the approaches of \cite{eynard2016coupled} and \cite{nogneng2018improved}, as it incorporates functional map coupling (via energy $E_1$) and descriptor commutativity (via $E_4$). Unlike our technique, however, it operates on fixed descriptor functions, and uses descriptor preservation constraints with the original and noisy descriptors.

For fairness of comparison, we used SHOT descriptors \cite{shot} as input to all methods, except BCICP \cite{ren2018continuous}, which uses carefully curated WKS \cite{aubry11} descriptors. Furthermore, we consider the results of FMNet \cite{litany2017deep} before and after applying the PMF-based post-processing as suggested in the original article. We also report results with ICP post-processing introduced in \cite{ovsjanikov2012functional}. Besides the accuracy plots shown in Figures~\ref{fig:sup-plot} and \ref{fig:unsup-plot}, we also include statistics such as maximum and 95th percentile in supplementary material.

\subsection{Evaluation and Results}

Figure~\ref{fig:sup-plot} summarizes the accuracy obtained by supervised methods on the three datasets whereas Figure~\ref{fig:unsup-plot} compares with unsupervised methods, using the evaluation~protocol introduced in \cite{kim11}. Note that in all cases, our network \textbf{SURFMNet}, (Ours-all), when optimized on all shapes achieves the best results even compared to the recent state-of-the-art method in \cite{ren2018continuous}. Furthermore, our method is comparable even to supervised learning techniques, GCNN \cite{boscaini2015learning} and FMNet \cite{litany2017deep} despite being purely unsupervised.

\begin{figure*}[t!]
\begin{center}
\begin{subfigure}{.11\paperwidth}
  \centering
  \includegraphics[trim={8cm 4cm 6cm 0}, clip, width=\linewidth]{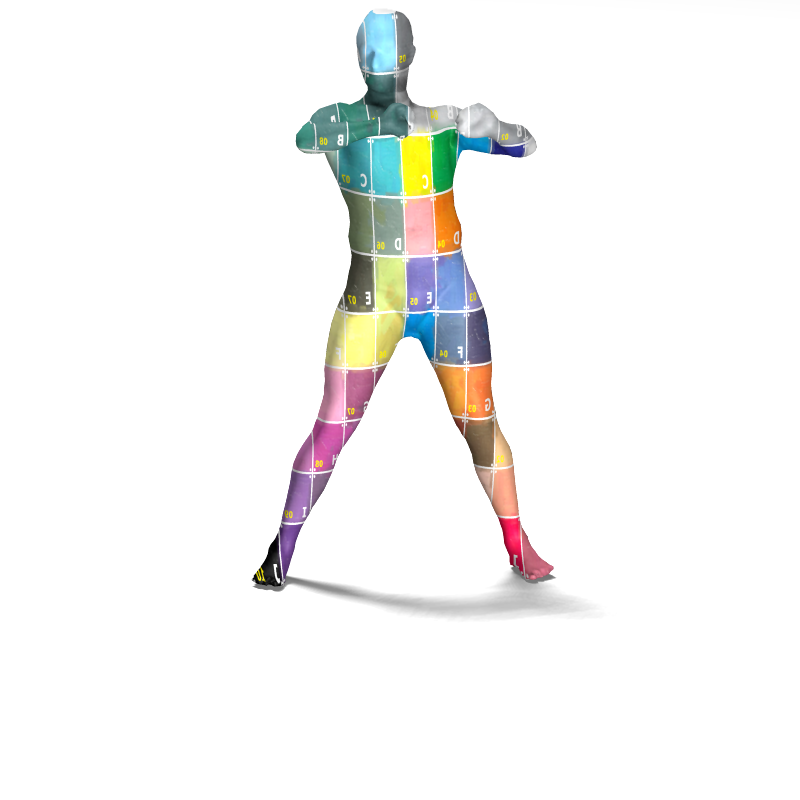}
  \caption*{Source}
  \label{fig:sfig1}
\end{subfigure}%
\begin{subfigure}{.11\paperwidth}
  \centering
  \includegraphics[trim={8cm 4cm 6cm 0}, clip, width = \linewidth]{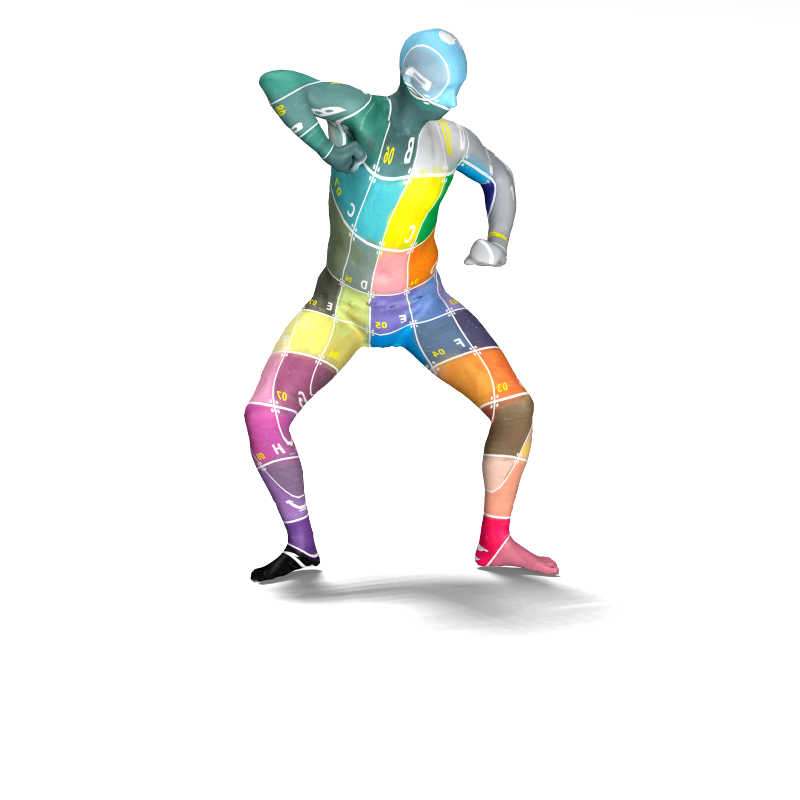}
  \caption*{Ground-Truth}
  \label{fig:sfig2}
\end{subfigure}
\begin{subfigure}{.11\paperwidth}
  \centering
  \includegraphics[trim={8cm 4cm 6cm 0}, clip, width = \linewidth]{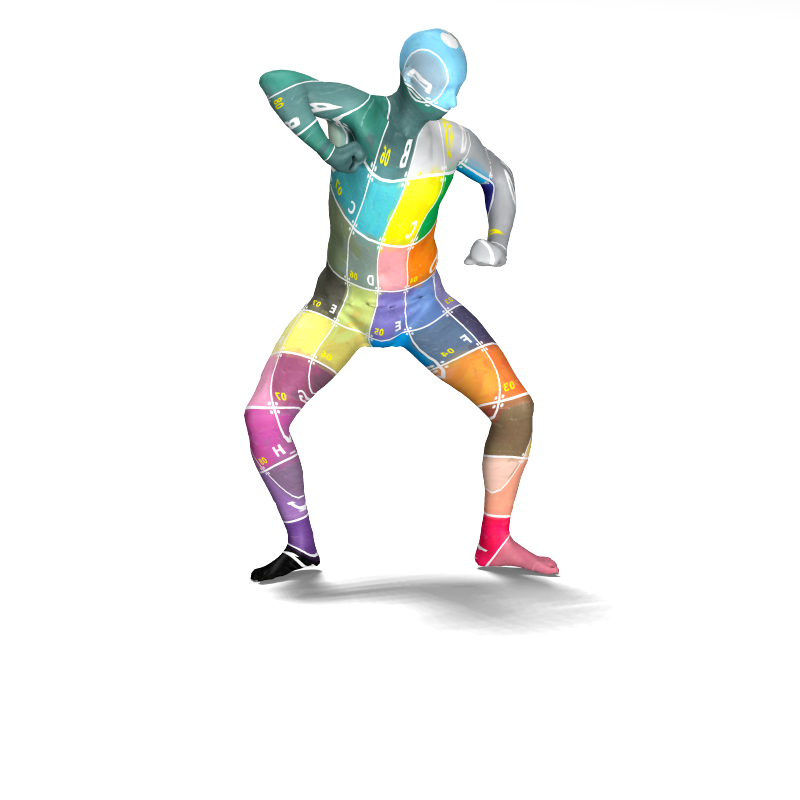}
  \caption*{SURFMNet}
  \label{fig:sfig2}
\end{subfigure}
\begin{subfigure}{.11\paperwidth}
  \centering
  \includegraphics[trim={8cm 4cm 6cm 0}, clip, width = \linewidth]{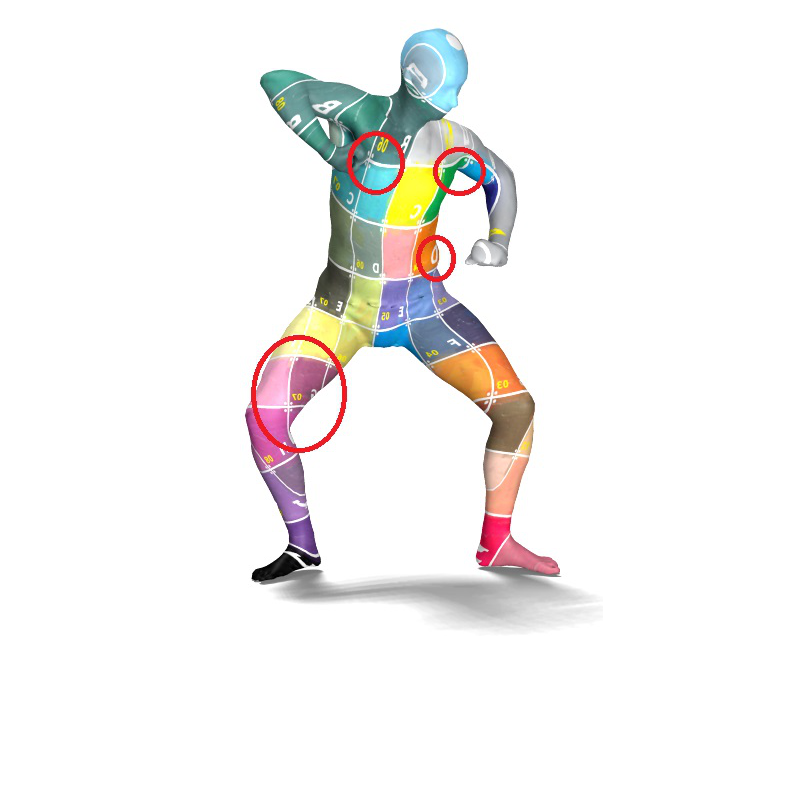}
  \caption*{BCICP}
  \label{fig:sfig2}
\end{subfigure}
\begin{subfigure}{.11\paperwidth}
  \centering
  \includegraphics[trim={8cm 4cm 6cm 0}, clip, width = \linewidth]{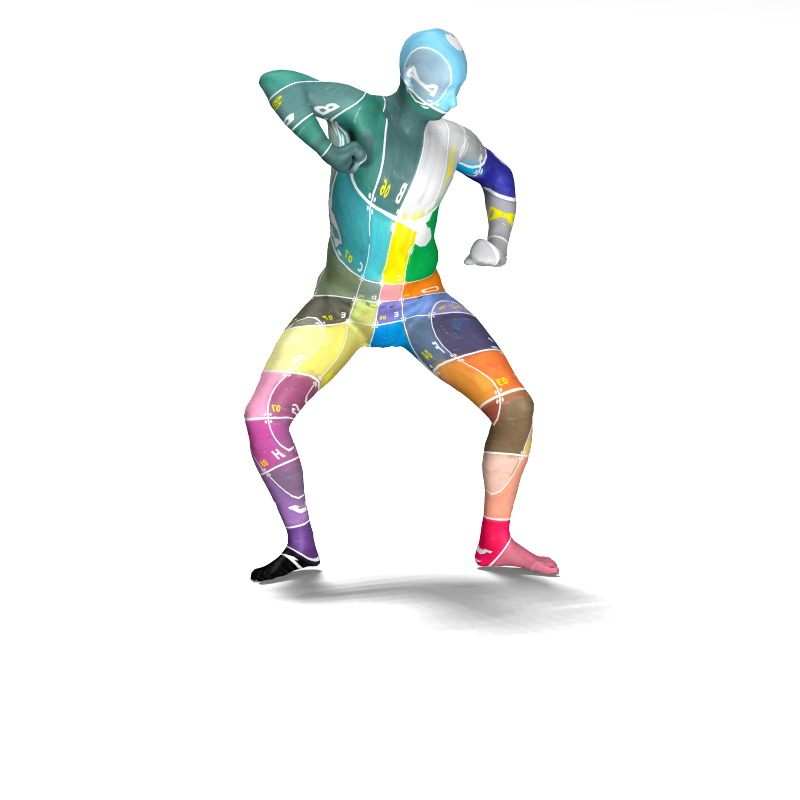}
  \caption*{PMF (heat)}
  \label{fig:sfig2}
\end{subfigure}
\begin{subfigure}{.11\paperwidth}
  \centering
  \includegraphics[trim={8cm 4cm 6cm 0}, clip, width = \linewidth]{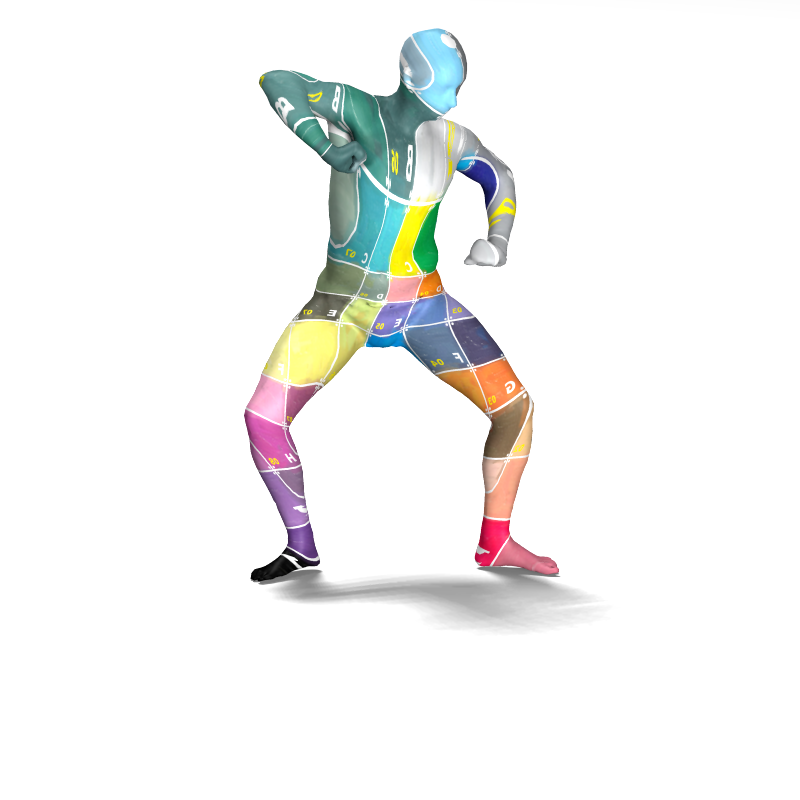}
  \caption*{PMF (gauss)}
  \label{fig:sfig2}
\end{subfigure}
\end{center}
\vspace{-4mm}
   \caption{Comparison of our method with \textit{Unsupervised} methods for texture transfer on the SCAPE remeshed dataset. Note that BCICP is roughly 7 times slower than our method and its shortcomings are marked with red circles.
   \vspace{-2mm}}
\label{fig:SCAPE2}
\end{figure*}

\begin{figure*}
\begin{center}
\begin{subfigure}{.1\paperwidth}
  \centering
  \includegraphics[trim={5cm 4cm 6cm 0}, clip, width=\linewidth]{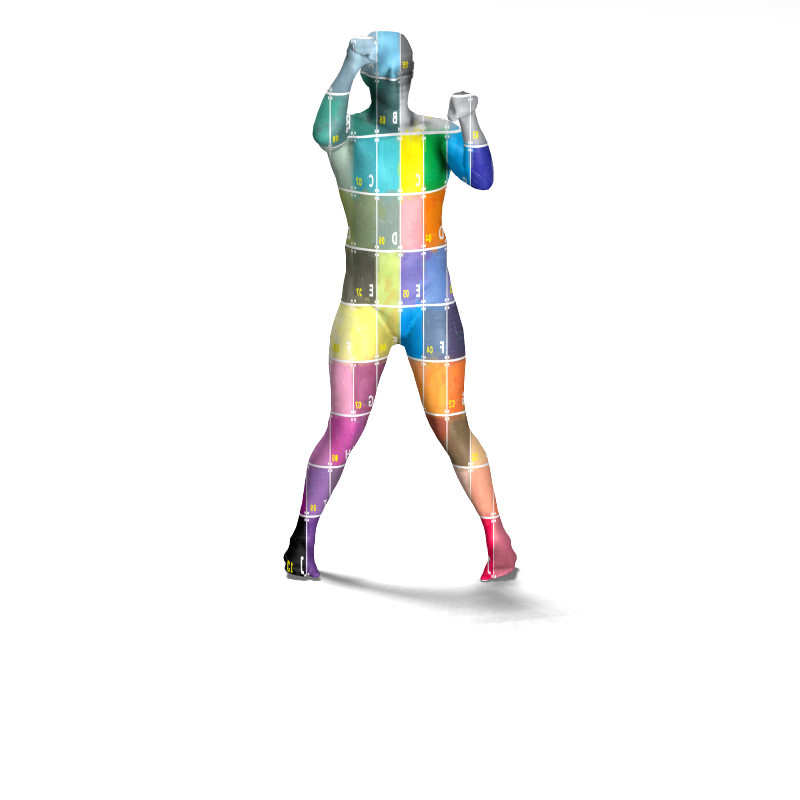}
  \caption*{Source}
  \label{fig:sfig1}
\end{subfigure}%
\begin{subfigure}{.1\paperwidth}
  \centering
  \includegraphics[trim={5cm 4cm 6cm 0}, clip, width = \linewidth]{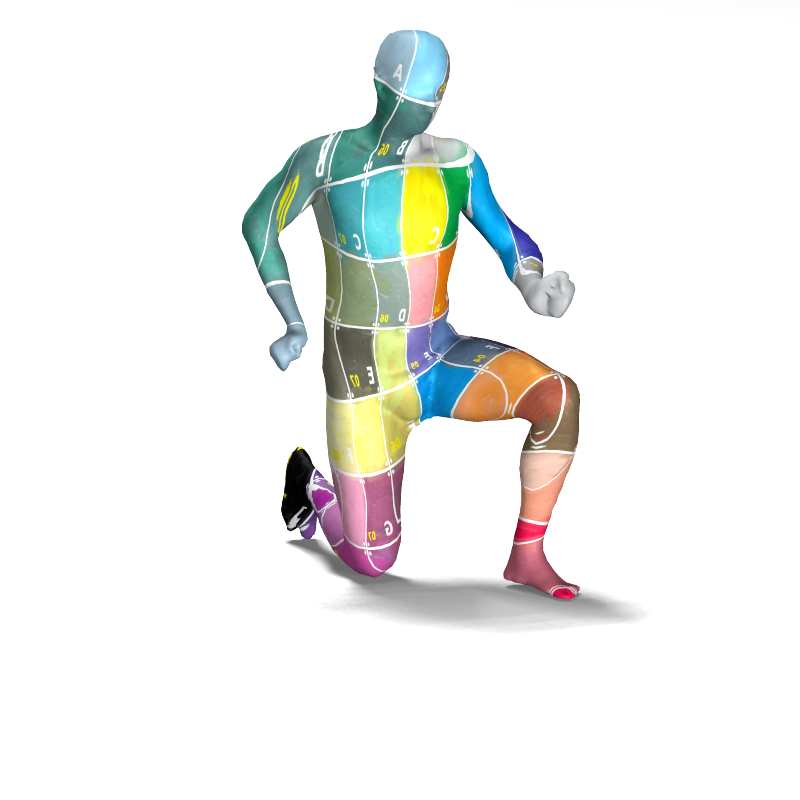}
  \caption*{Ground-Truth}
  \label{fig:sfig2}
\end{subfigure}
\begin{subfigure}{.1\paperwidth}
  \centering
  \includegraphics[trim={5cm 4cm 6cm 0}, clip, width = \linewidth]{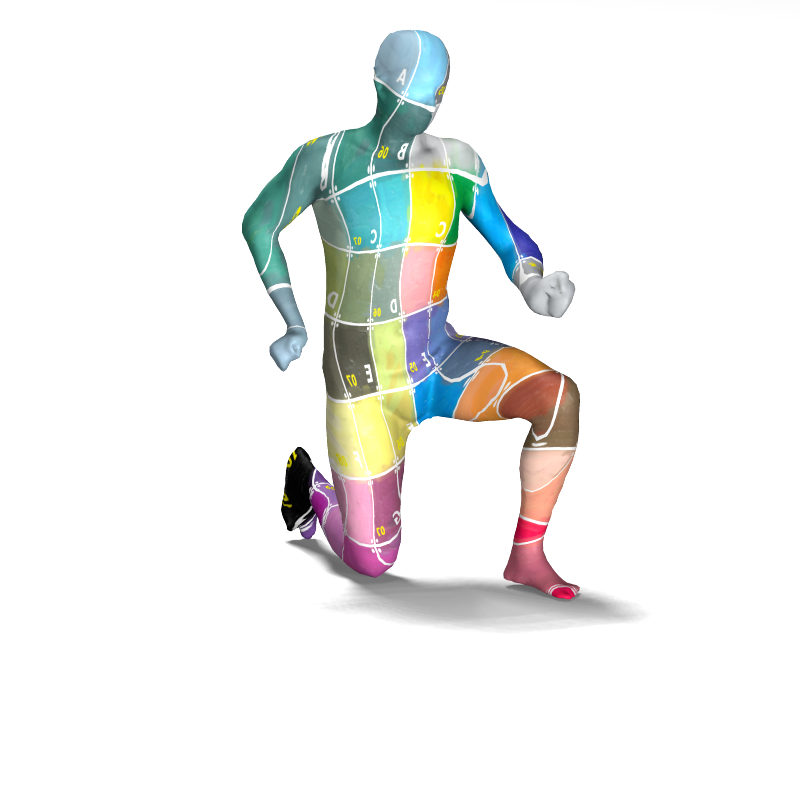}
  \caption*{SURFMNet$+$ICP}
  \label{fig:sfig2}
\end{subfigure}
\begin{subfigure}{.1\paperwidth}
  \centering
  \includegraphics[trim={5cm 4cm 6cm 0}, clip, width = \linewidth]{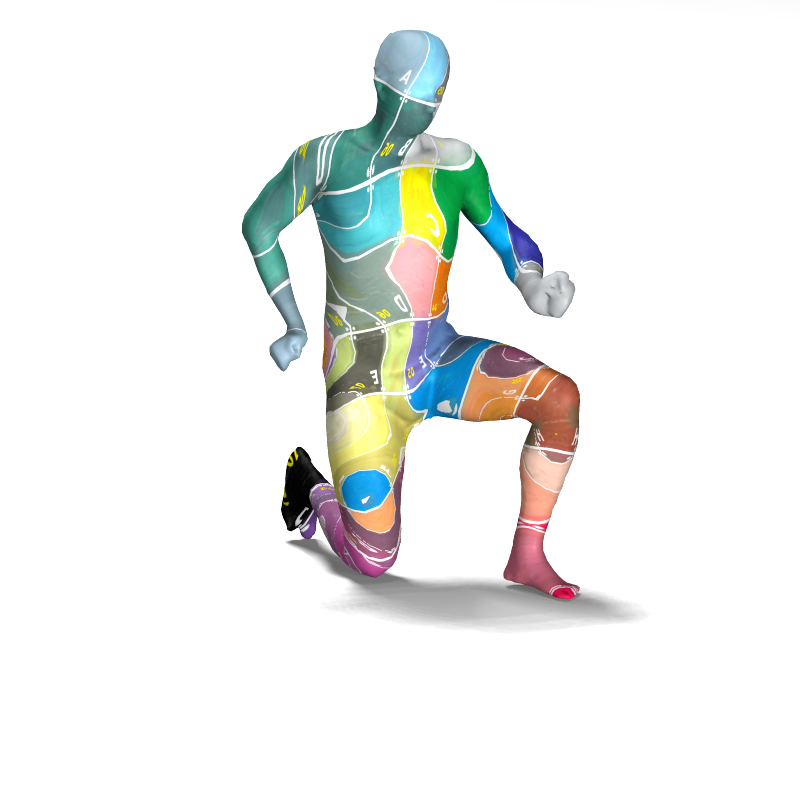}
  \caption*{SURFMNet}
  \label{fig:sfig2}
\end{subfigure}
\begin{subfigure}{.1\paperwidth}
  \centering
  \includegraphics[trim={5cm 4cm 6cm 0}, clip, width = \linewidth]{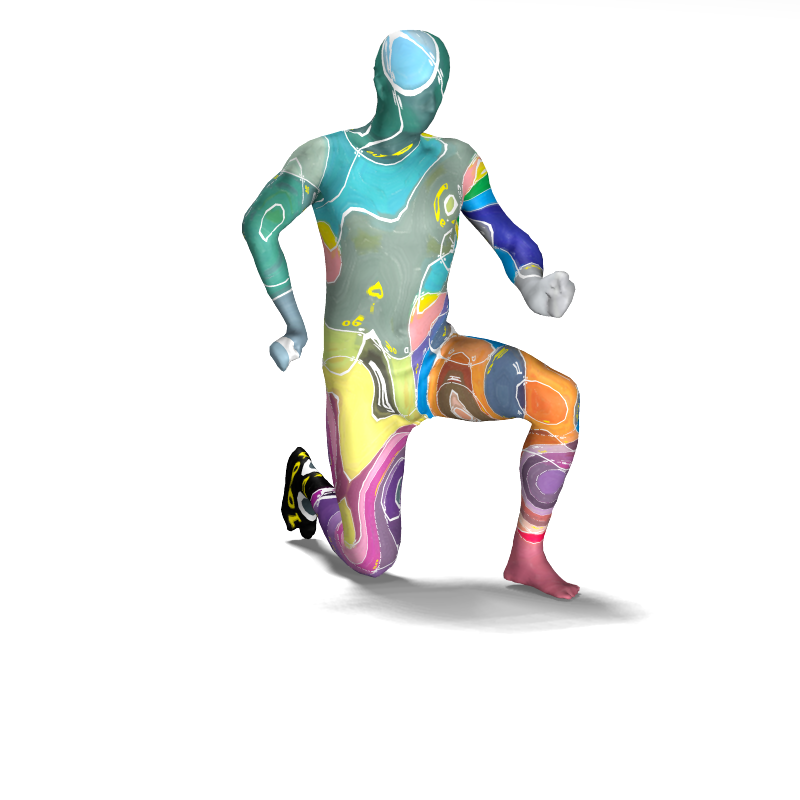}
  \caption*{FMNet}
  \label{fig:sfig2}
\end{subfigure}
\begin{subfigure}{.1\paperwidth}
  \centering
  \includegraphics[trim={5cm 4cm 6cm 0}, clip, width = \linewidth]{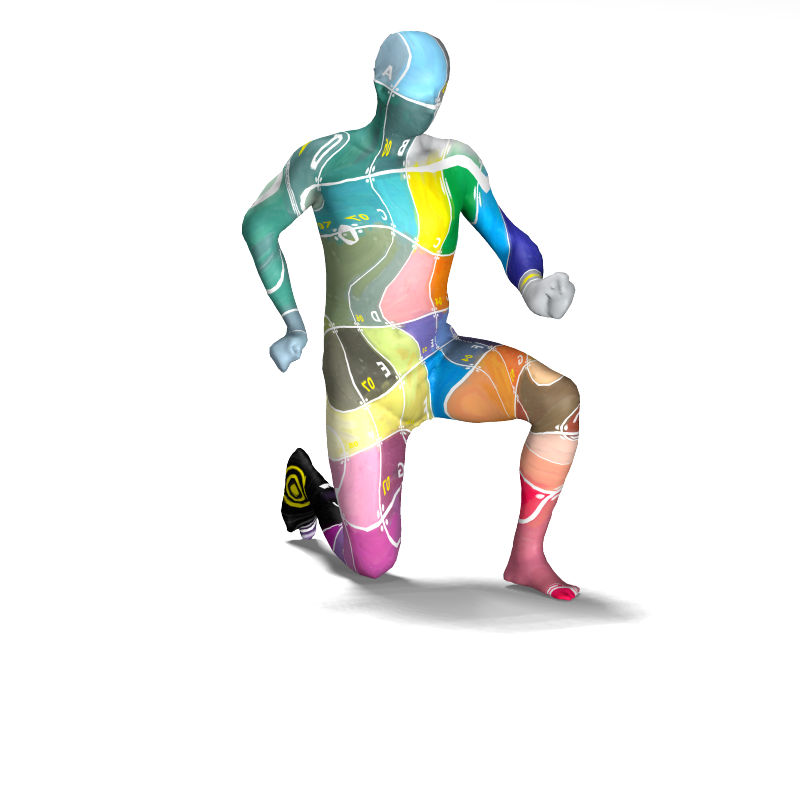}
  \caption*{FMNet $+$ PMF}
  \label{fig:sfig2}
\end{subfigure}
\begin{subfigure}{.1\paperwidth}
  \centering
  \includegraphics[trim={5cm 4cm 6cm 0}, clip, width = \linewidth]{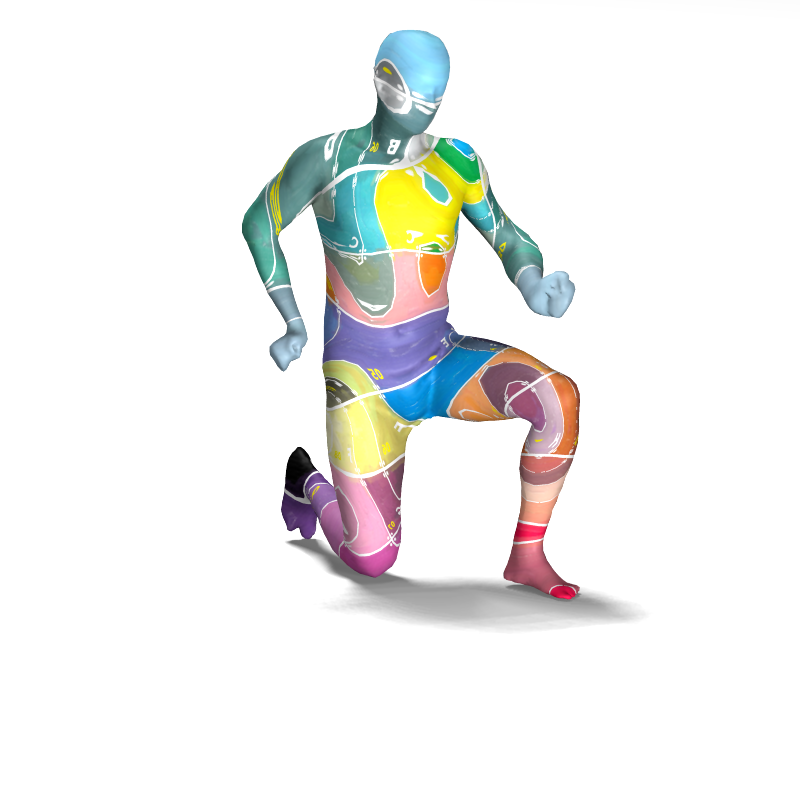}
  \caption*{GCNN}
  \label{fig:sfig2}
\end{subfigure}
\end{center}
\vspace{-4mm}
  \caption{Comparison of our method with \textit{Supervised} method for texture transfer on the SCAPE remeshed dataset.
  \vspace{-1mm}}
\label{fig:sup}
\end{figure*}

Remark that the remeshed datasets are significantly harder for both supervised and unsupervised methods, since the shapes are no longer identically meshed and in 1-1 correspondence. We have observed this difficulty also while training supervised FMNet and GCNN techniques with very slow  convergence during training. On both of these datasets, our approach achieves the lowest average error, reported in Figure~\ref{fig:sup-plot} and \ref{fig:unsup-plot}. Note that on the remeshed FAUST dataset, as shown in Figure~\ref{fig:sup-plot}, only GCNN \cite{boscaini2015learning} produces a similarly large fraction of correspondences with a small error. However, this method is \emph{supervised}. On the remeshed SCAPE dataset, our method leads to the best results across all measures, despite being purely unsupervised. 

\paragraph{Postprocessing Results}  As shown in Figures~\ref{fig:sup-plot} and \ref{fig:unsup-plot} our method can often obtain high quality results even without any post-processing. Nevertheless, in the challenging cases such as the SCAPE remeshed dataset, when trained on a subset of shapes, it can also benefit from an efficient ICP-based refinement. This refinement, does not require computing geodesic distances and does not require the shapes to have the same number of points, thus maintaining the flexibility and efficiency of our pipeline. 

\paragraph{Correlation with actual Geodesic loss} We further
investigated if there is a correlation between the value of our loss
and the quality of correspondence. Specifically, whether
minimizing our loss function, mainly consisting of regularization
terms on estimated functional maps, corresponds to minimizing the
geodesic loss with respect to the unknown ground truth map. We found
strong correlation between the two and share a plot in the
supplementary material.

\begin{table}[t!]
\begin{center}
\resizebox{0.48\textwidth}{!}{\begin{tabular}{@{}lcccc|c@{}}
\toprule
& \multicolumn{5}{c}{Runtime}  \\ \cmidrule(l){2-6} 
\textbf{Methods} & Pre-processing       & Training                & Testing            & Post-processing      & Total \\ \cmidrule(r){1-6}
FMNet    &  60s  & 1500s  & 0.3s & N/As & 1650s  \\
FMNet + PMF   &  60s  & 1500s  & 0.3s & 30s & 1680s  \\
Fmap Basic  & 10s & N/A & 60s & N/A &  120s \\
BCICP   & N/A & N/A & 60s & 180s & 240s\\
SURFMNet & 10s &   25s &  0.3s  & N/A & \textbf{35s}\\ 
SURFMNet + ICP & 10s &   25s &  0.3s  & 10s & 45s\\ \bottomrule
\end{tabular}}
\caption{Runtime of different methods averaged over 190 shape pairs.
\label{tab:runtime}
\vspace{-9mm}}
\end{center}
\end{table}

\paragraph{Qualitative and Runtime Comparison} Figures~\ref{fig:SCAPE2} and \ref{fig:sup} show examples shape pairs and maps obtained between them using different methods, visualized via texture transfer. Note the continuity and quality of the maps obtained using our method, compared to other techniques (more results in supplementary material). One further advantage of our method is its efficiency, since we do not rely on the computation of geodesic matrices and operate entirely in the spectral domain. 
Table~\ref{tab:runtime} compares the  run-time of the best performing methods on an Intel Xeon 2.10GHz machine with an NVIDIA Titan X GPU. Note that our method is over an order of magnitude faster than FMNet and significantly faster than the currently best unsupervised BCICP.

\section{Conclusion \& Future Work}
We presented an unsupervised method for computing correspondences between shapes. Key to our approach is a bi-level optimization formulation, aimed to optimize descriptor functions, while promoting the structural properties of the entire map, obtained from them via the functional maps framework. 
Remarkably, our approach achieves similar, and in some cases superior performance even to supervised correspondence techniques.

In the future, we plan to incorporate other penalties on functional maps, e.g., those arising from recently-proposed kernalization approaches \cite{wang2018kernel}, or for promoting orientation preserving maps\cite{ren2018continuous} and also incorporate cycle consistency constraints \cite{huang2014functional}. Finally, it would  be interesting to extend our method to partial and non-isometric shapes and matching other modalities, such as images or point clouds, since it opens the door to linking the properties of local descriptors to global map consistency.

\noindent\textbf{Acknowledgements} Parts of this work were supported by the ERC Starting Grant StG-2017-758800 (EXPROTEA), KAUST OSR Award No. CRG-2017-3426, and a gift from Nvidia. We are grateful to \textbf{Jing Ren}, Or Litany, Emanuele Rodol{\`a} and Adrien Poulenard for their help in performing quantitative comparisons and producing qualitative results.

\bibliographystyle{ieee}
\bibliography{sec-bib}

\section{Supplement}

\renewcommand{\thesubsection}{\Alph{subsection}}

\subsection{Correlation with actual geodesic loss}\label{sup:corr} 
\begin{figure}[!h]
  \centering
  \includegraphics[scale = 0.42]{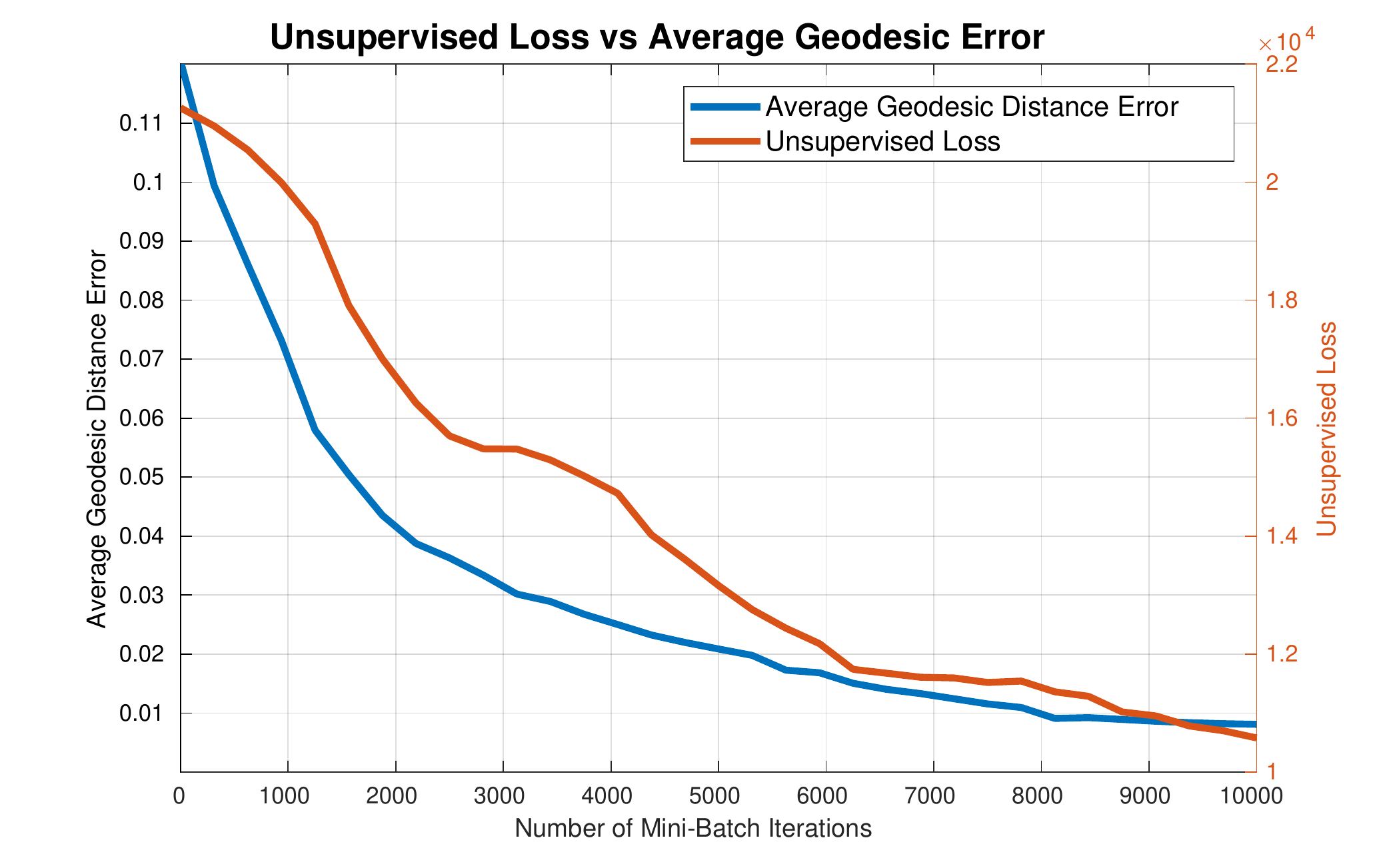}
  \caption{Correlation with average geodesic loss computed from ground truth correspondences.}
  \label{fig:corr}
\end{figure}

To support the claim made in the subsection 'Evaluation and Results'
, we include a plot here to visualize the correlation between our loss and the actual geodesic loss. As evident in Figure \ref{fig:corr}, there is a strong correlation between our loss value and the quality of correspondence as measured by average geodesic error.

\subsection{Detailed Tabular Quantitative Comparison}\label{sup:tab} 
\begin{figure*}[t!]
\begin{center}
\begin{subfigure}{.265\paperwidth}
  \centering
  \includegraphics[trim={.5cm 0 1.5cm 0}, clip, width=\linewidth]{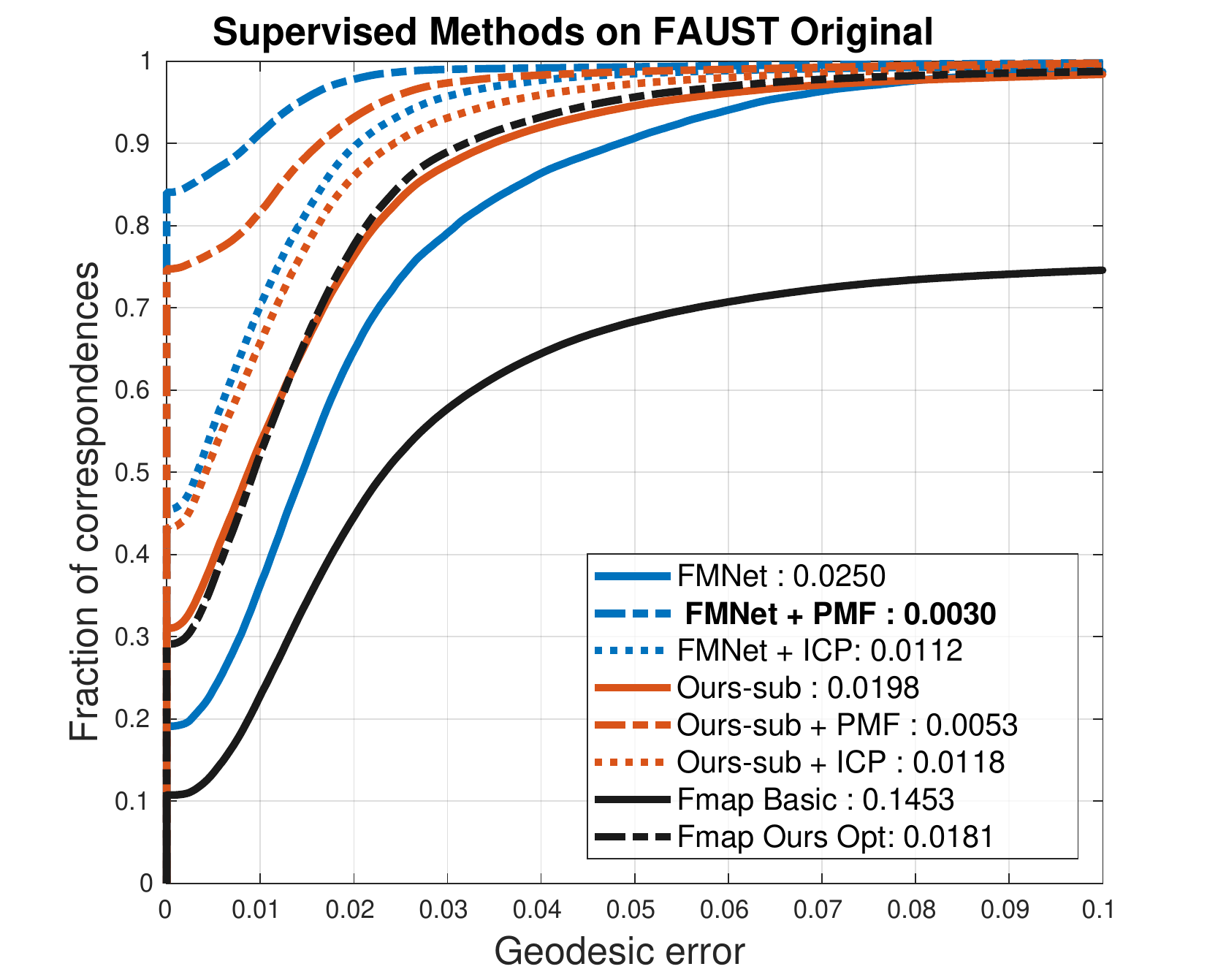}
  \label{fig:sfig1}
\end{subfigure}%
\begin{subfigure}{.265\paperwidth}
  \centering
  \includegraphics[trim={.5cm 0 1.5cm 0}, clip, width = \linewidth]{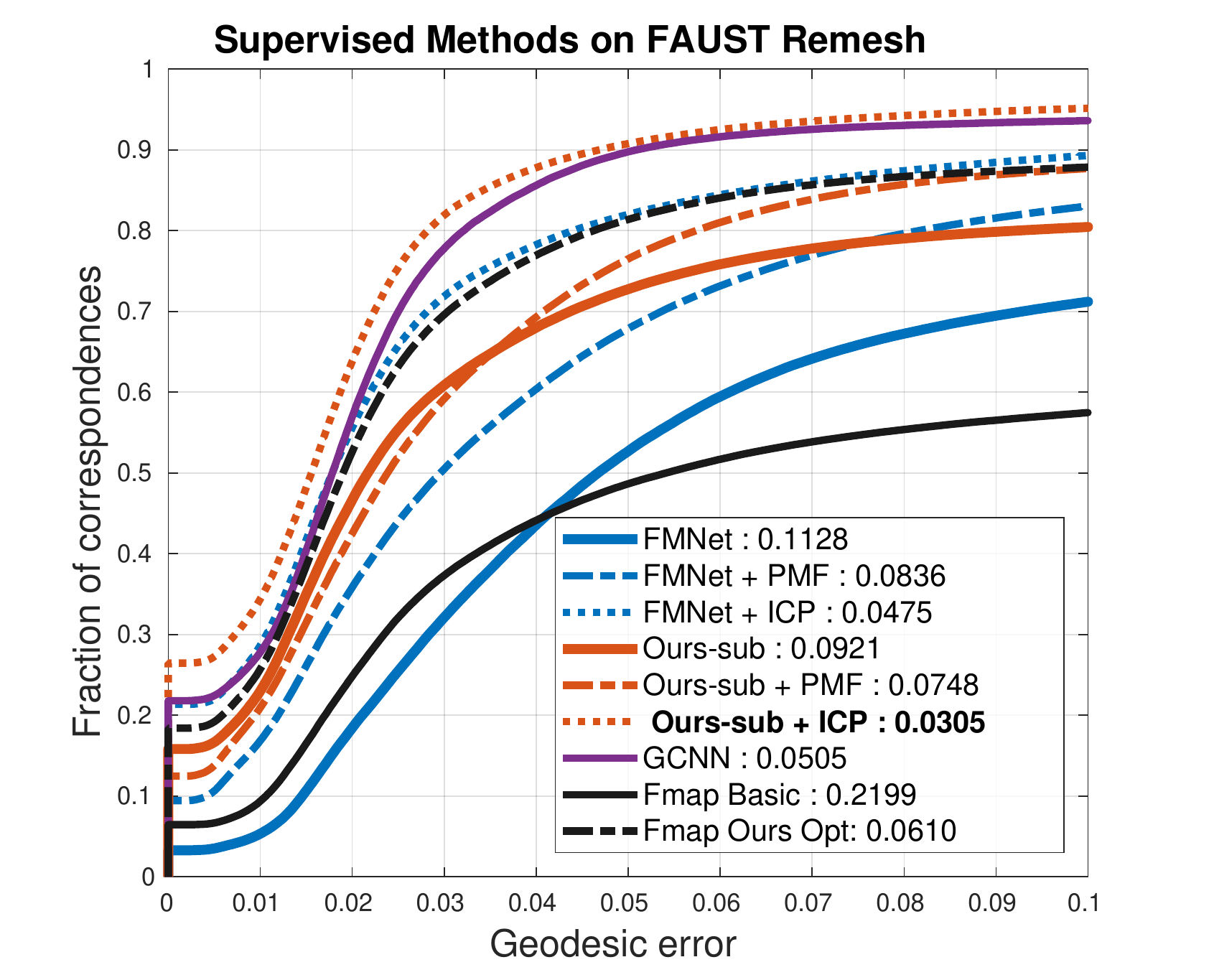}
  \label{fig:sfig2}
\end{subfigure}
\begin{subfigure}{.265\paperwidth}
  \centering
  \includegraphics[trim={.5cm 0 1.5cm 0}, clip, width = \linewidth]{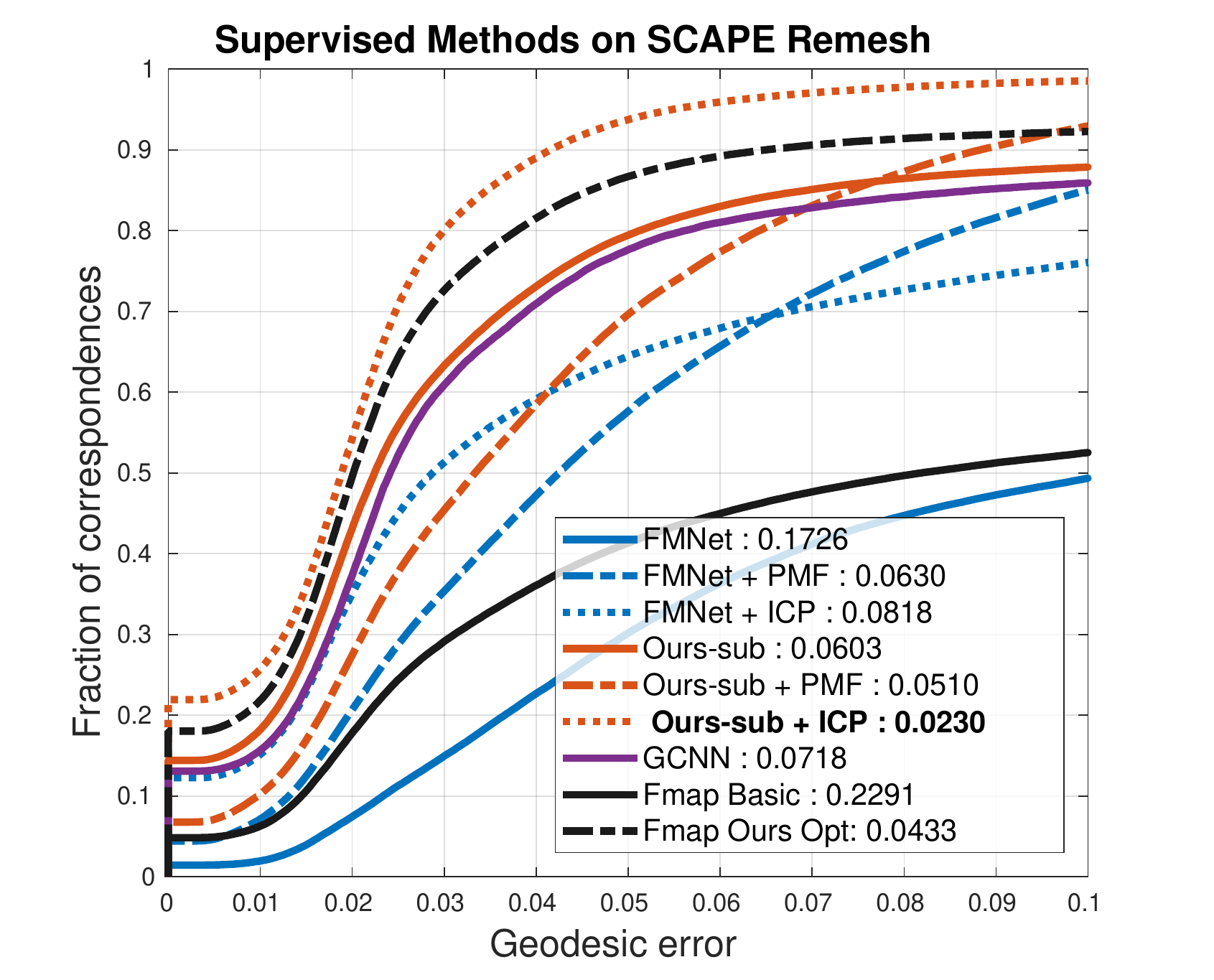}
  \label{fig:sfig2}
\end{subfigure}
\end{center}
\vspace{-8mm}
   \caption{Quantitative evaluation of pointwise correspondences comparing our method with Supervised Methods. 
   \vspace{-2mm}}
\label{fig:sup-plot_sup}
\end{figure*}

\begin{figure*}[t!]
\begin{center}
\begin{subfigure}{.27\paperwidth}
  \centering
  \includegraphics[trim={1cm 0 1cm 0}, clip, width=1.05\linewidth]{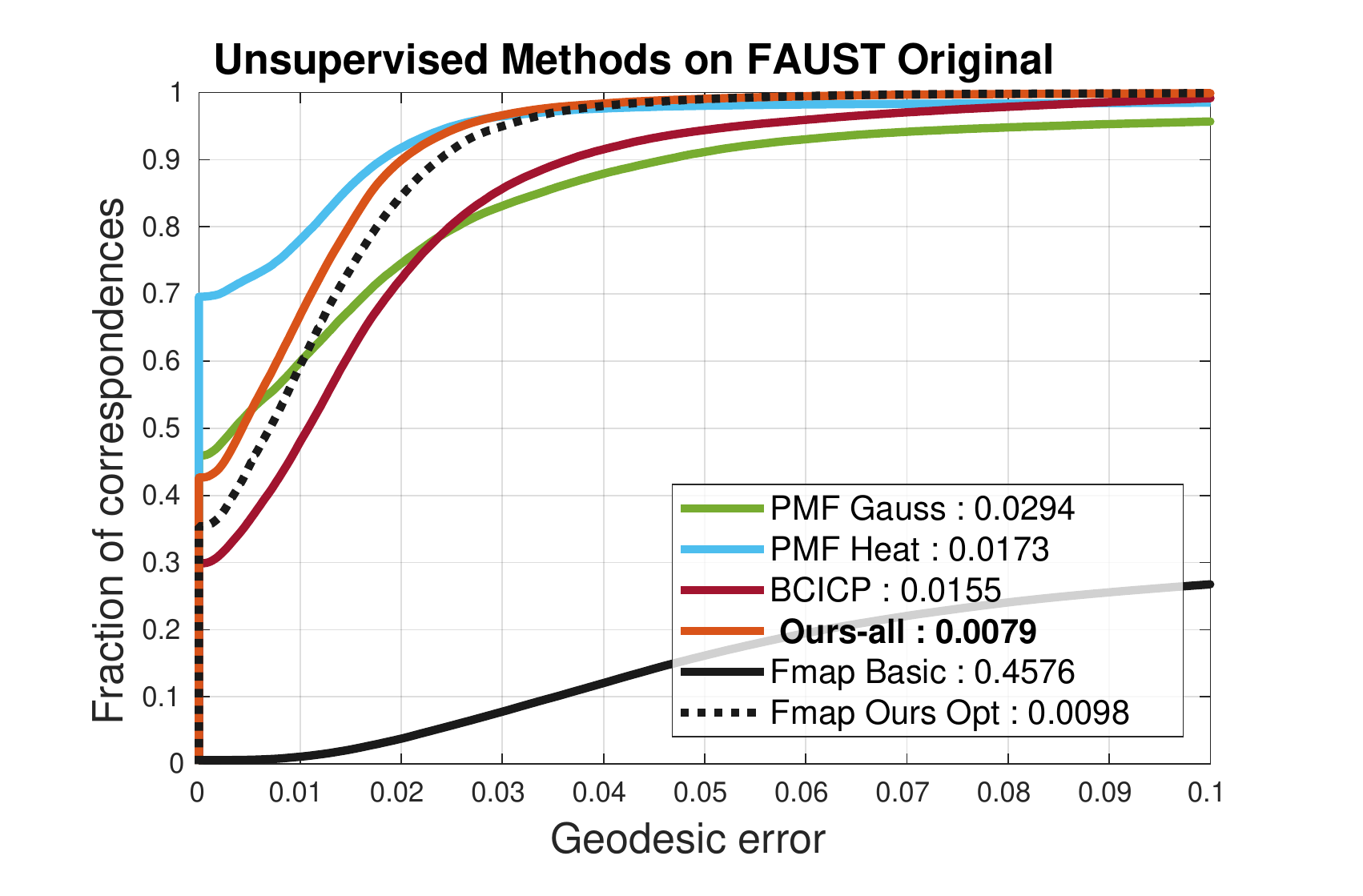}
  \label{fig:sfig1}
\end{subfigure}%
\begin{subfigure}{.265\paperwidth}
  \centering
  \includegraphics[trim={.5cm 0 1cm 0}, clip, width = \linewidth]{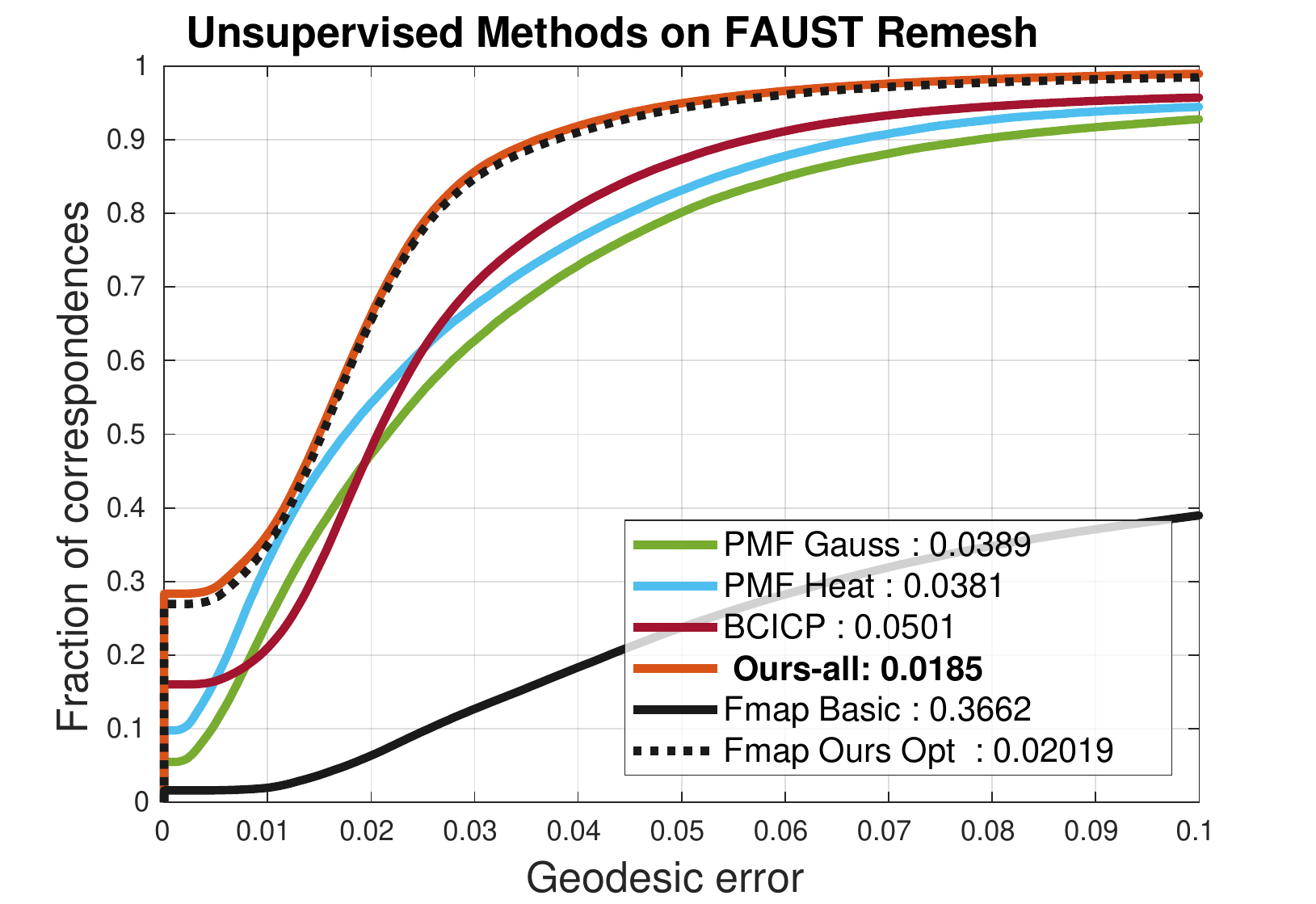}
  \label{fig:sfig2}
\end{subfigure}
\begin{subfigure}{.265\paperwidth}
  \centering
  \includegraphics[trim={.5cm 0 1.5cm 0}, clip, width = \linewidth]{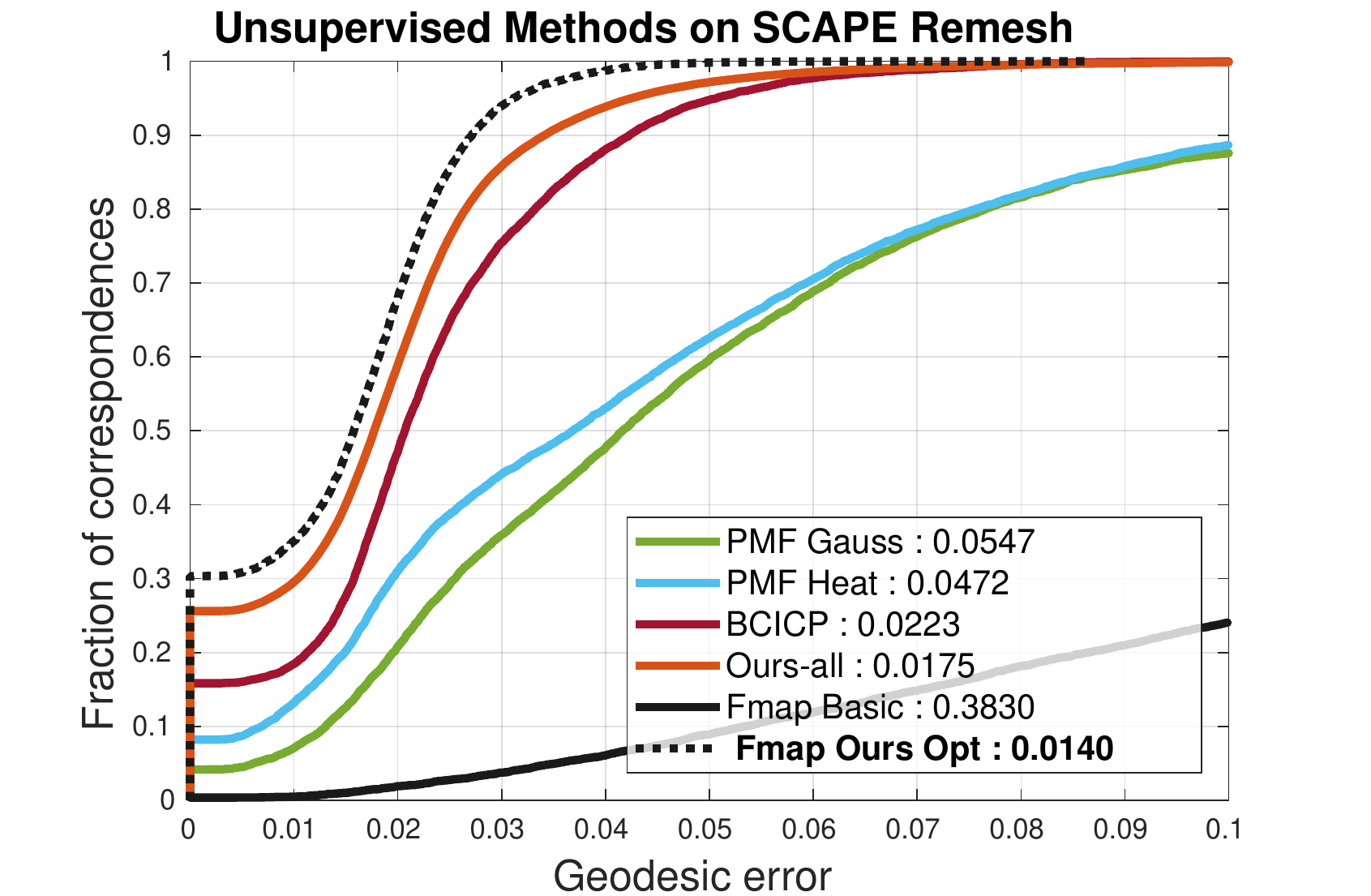}
  \label{fig:sfig2}
\end{subfigure}
\end{center}
\vspace{-8mm}
   \caption{Quantitative evaluation of pointwise correspondences comparing our method with Unsupervised Methods. 
   \vspace{-2mm}}
\label{fig:unsup-plot_sup}
\end{figure*}

Besides the average geodesic error reported for quantitative comparison in Figures $3$ and $4$, we provide detailed statistics in Table \ref{results_all}. Note that Table \ref{results_all} also includes 'Fmap Ours Opt' which is equivalent to ``Fmap Basic'' but uses the learned descriptors instead of original ones. Its competitive performance across all datasets proves quantitatively the utility of learning descriptors. Figures \ref{fig:sup-plot_sup} and \ref{fig:unsup-plot_sup} illustrate this further. For completeness, in Table \ref{tab:ablation_sup}, we also provide a detailed ablation study with different combinations of penalties. 

 \begin{table*}[t!]
 \begin{center}
 \resizebox{\textwidth}{!}{\begin{tabular}{l|ccc|ccc|ccc}
 \toprule
 (Results are $\times 10^{-3}$) & \multicolumn{3}{c}{FAUST 7k}     & \multicolumn{3}{c}{FAUST 5k}     & \multicolumn{3}{c}{SCAPE 5k}     \\
 \textbf{Supervised Methods}    & Mean & 95th Percentile & Maximum & Mean & 95th Percentile & Maximum & Mean & 95th Percentile & Maximum \\
 FMNet                 &    25.01  & 63.11 &  1207.8       &  112.8   & 451.8 & 1280.6  &172.6&543.8   &1399.6\\
SURFMNet Subset          &   19.83   & 52.11 & 1204.0        &  92.09  & 493.6 & 1279.4   &60.32 & 329.8& 1068.7\\
 FMNet + PMF          & \textbf{2.98} & 14.10   & 1222.7 & 83.61  & 395.7 & 1576.4  & 63.00 &159.8 & 1561.5        \\
 SURFMNet-sub + PMF &    5.33  &  22.90 & 1302.4 & 74.80 & 408.5 & 1619.3 & 51.03 &111.5   & 1555.6\\
 FMNet + ICP           & 11.16 & 27.91 & 1206.8 & 47.53 & 237.3  &   1348.6  &  81.76& 341.4  &1226.5   \\
 SURFMNet-sub + ICP & 11.79 & 35.76 & 1088.4 & \textbf{30.47} & 95.64   &  1277.3   & \textbf{23.00} &54.76 & 73.18         \\
 GCNN                  &   \_  &  \_ & \_ &   50.49   & 206.3  & 1578.2 &71.85& 374.2 & 1523.7\\
 \textbf{Unsupervised Methods}    &  &   &  &  &   &  &  &   &  \\

 BCICP                 & 15.46  & 53.27   &  572.4   &    31.08  &    64.51 &  1149.9   & 22.28 &   50.60  &107.5 \\
 PMF (Gaussian Kernel) & 29.42 &  83.80  & 1168.1 & 75.13     & 236.9   & 1632.7 &  54.68 & 156.9 & 465.1 \\
 PMF (Heat Kernel)     & 17.26 &  25.06 & 1168.1 &  31.08    & 64.51 & 1150.0 &  47.23 & 133.4 & 802.1  \\
 Fmap Basic            &   457.56   &  1171.4  & 1568.4 &   366.2   & 1159.0 & 1549.1 &383.0& 1043.7 & 1280.3 \\
 Fmap Ours Opt  &    9.75 & 30.02 & 420.2 &  20.19& 53.24& 1169.5 & \textbf{13.98} & 31.16 & 86.45 \\
 SURFMNet-all             &  \textbf{7.89}  &  26.01 & 572.4  &   \textbf{18.56}   & 50.25 & 1156.3 & 17.50 & 42.50  & 228.8
 \end{tabular}}
 \end{center}
 \caption{Quantitative comparison on all three benchmark datasets for shape correspondence problem.}
 \label{results_all}
 \end{table*}
 
\begin{table*}
\begin{center}
\resizebox{\textwidth}{!}{\begin{tabular}{l|cccccccccccc}
\toprule
\textbf{Methods} & E1+E2+E3+E4 & E3   & E1+E2+E3 & E1+E3+E4 & E1   & E2+E3+E4 & E1+E2+E4 & E2   & E4 & FMNet &Ours-Sub  & Ours-all    \\ \midrule
\textbf{Mean Geodesic Error}& 0.044        & 0.073 & 0.081     & 0.077     & 0.111 & 0.079     & 0.126     & 0.135 & 0.330 & 0.025 & 0.020 & \textbf{0.008}
\end{tabular}}
\end{center}
 \caption{Ablation study of penalty terms in our method and comparison with the supervised FMNet
  on the FAUST benchmark.
\label{tab:ablation_sup}}
\end{table*}

\subsection{Sensitivity to number of basis functions}\label{sup:eig} 
\begin{figure}[!h]
  \centering
  \includegraphics[width=0.55\textwidth]{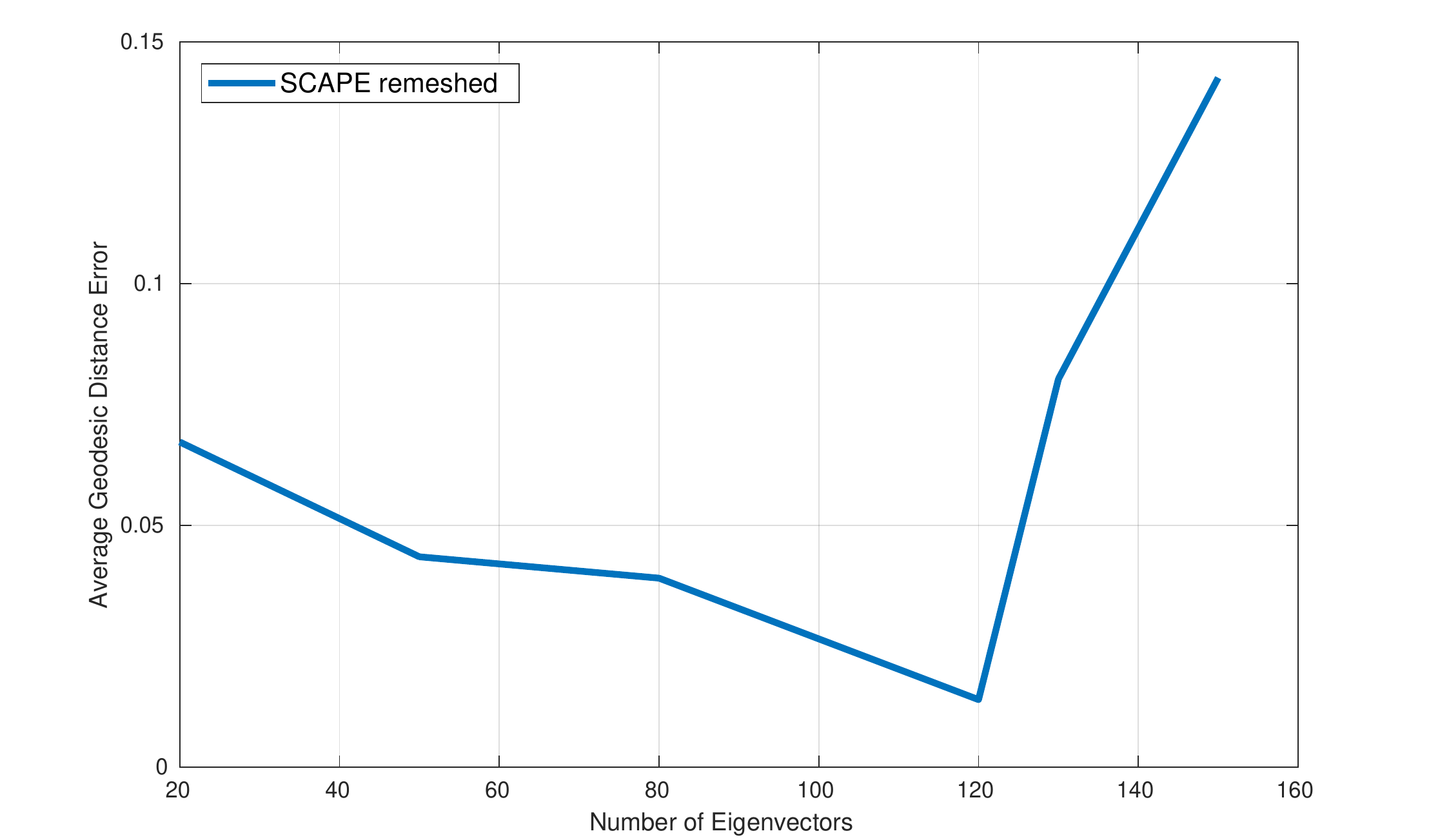}
  \caption{Accuracy of our method on the SCAPE remeshed dataset as the number of eigenfunctions is varied from 20 to 150.}
  \label{fig:eigen}
\end{figure}

Figure \ref{fig:eigen} shows the sensitivity of our network SURFMNet on the SCAPE remeshed dataset as the number of eigen functions are varied from 20 to 150. We train the network each time with 10000 mini batch steps. As evident, we obtain best result using 120. However, when trained on an individual dataset and tested on a different one, we see over-fitting when using a large eigen-basis. We attribute this phenomenon to the initialization of our descriptors with SHOT which is a very local descriptor and is not robust to very strong mesh variability. However, over-fitting is minimal when we train together on a relatively larger subset of SCAPE and FAUST and test on a different subset of shapes from both datasets, with smaller eigen basis.

\begin{figure*}[t!]
\begin{center}
\begin{subfigure}{.1\paperwidth}
  \centering
  \includegraphics[trim={9cm 4cm 8cm 0}, clip, width=\linewidth]{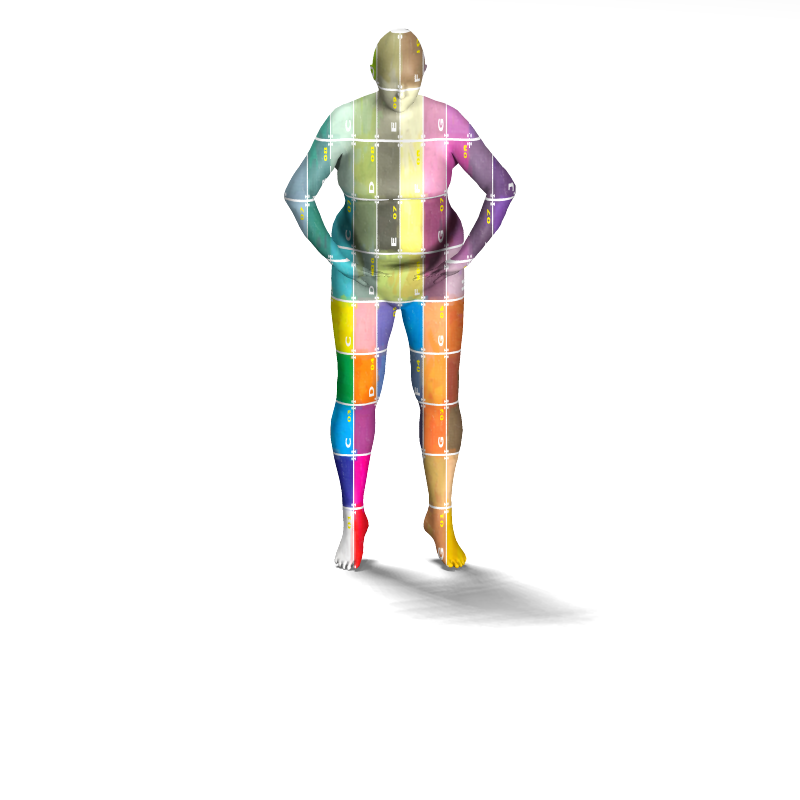}
  \caption*{Source}
  \label{fig:sfig1}
\end{subfigure}%
\begin{subfigure}{.1\paperwidth}
  \centering
  \includegraphics[trim={9cm 4cm 8cm 0}, clip, width = \linewidth]{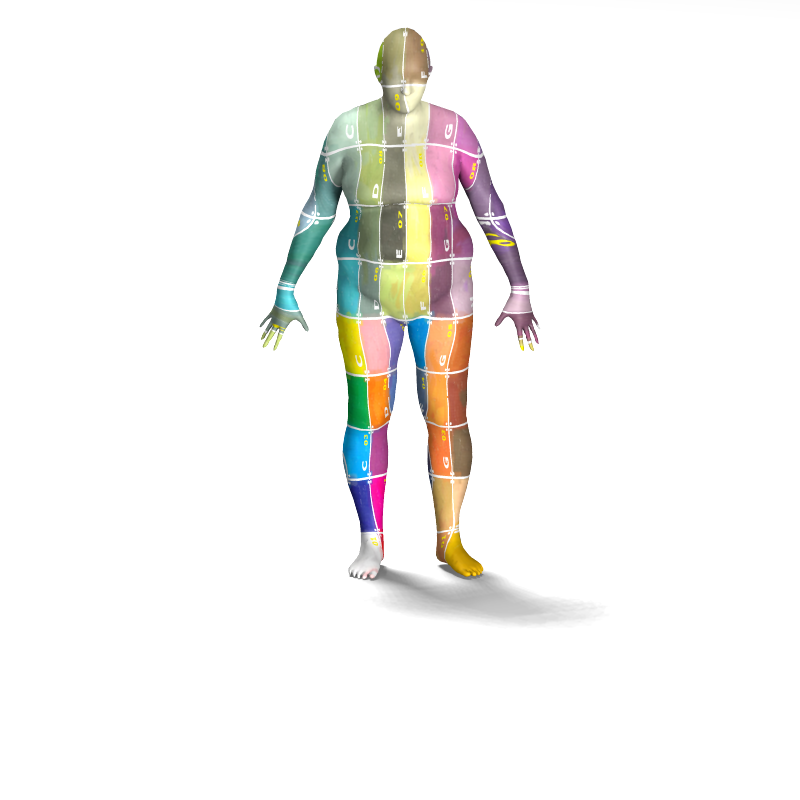}
  \caption*{Ground-Truth}
  \label{fig:sfig2}
\end{subfigure}
\begin{subfigure}{.1\paperwidth}
  \centering
  \includegraphics[trim={9cm 4cm 8cm 0}, clip, width = \linewidth]{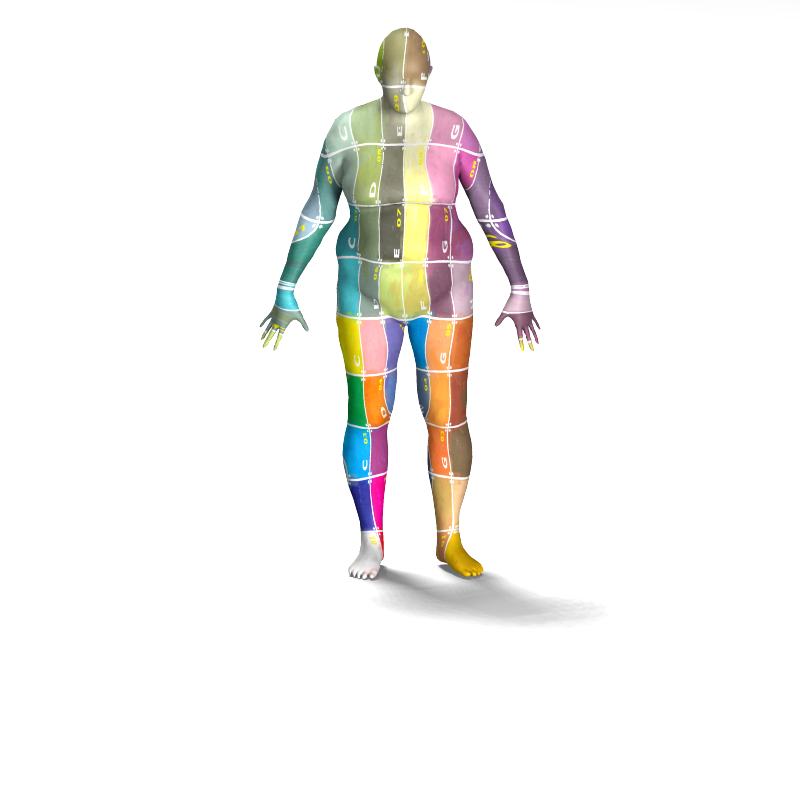}
  \caption*{SUFMNet + ICP}
  \label{fig:sfig2}
\end{subfigure}
\begin{subfigure}{.1\paperwidth}
  \centering
  \includegraphics[trim={9cm 4cm 8cm 0}, clip, width = \linewidth]{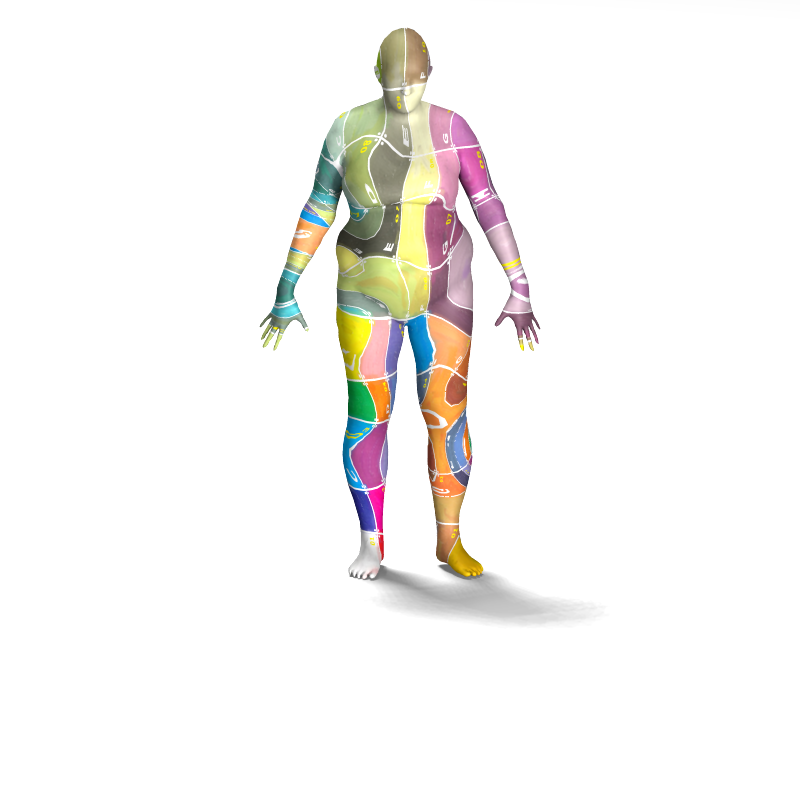}
  \caption*{SUFMNet}
  \label{fig:sfig2}
\end{subfigure}
\begin{subfigure}{.1\paperwidth}
  \centering
  \includegraphics[trim={9cm 4cm 8cm 0}, clip, width = \linewidth]{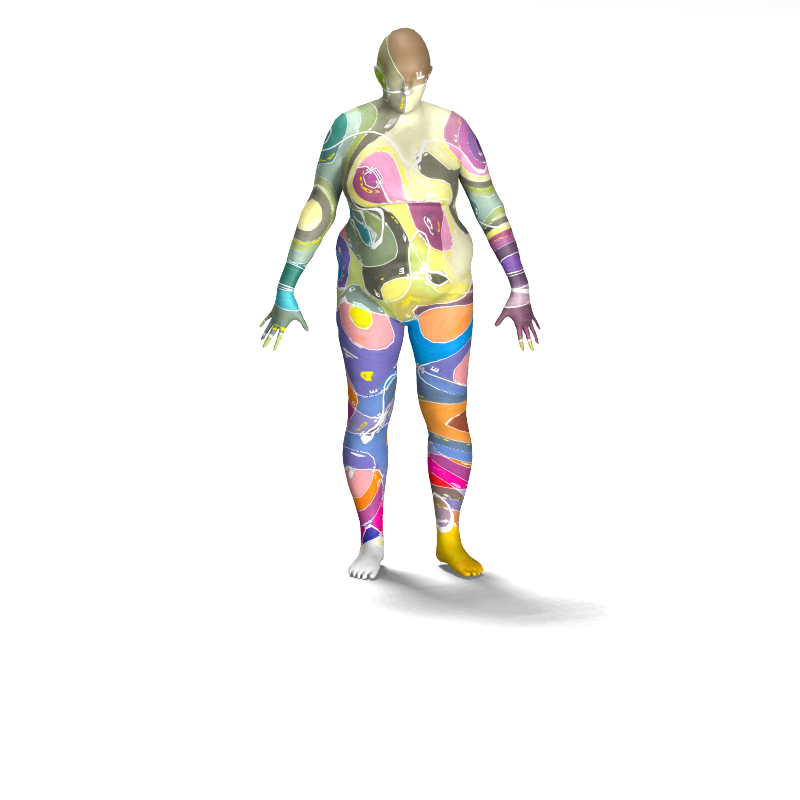}
  \caption*{FMNet}
  \label{fig:sfig2}
\end{subfigure}
\begin{subfigure}{.1\paperwidth}
  \centering
  \includegraphics[trim={9cm 4cm 8cm 0}, clip, width = \linewidth]{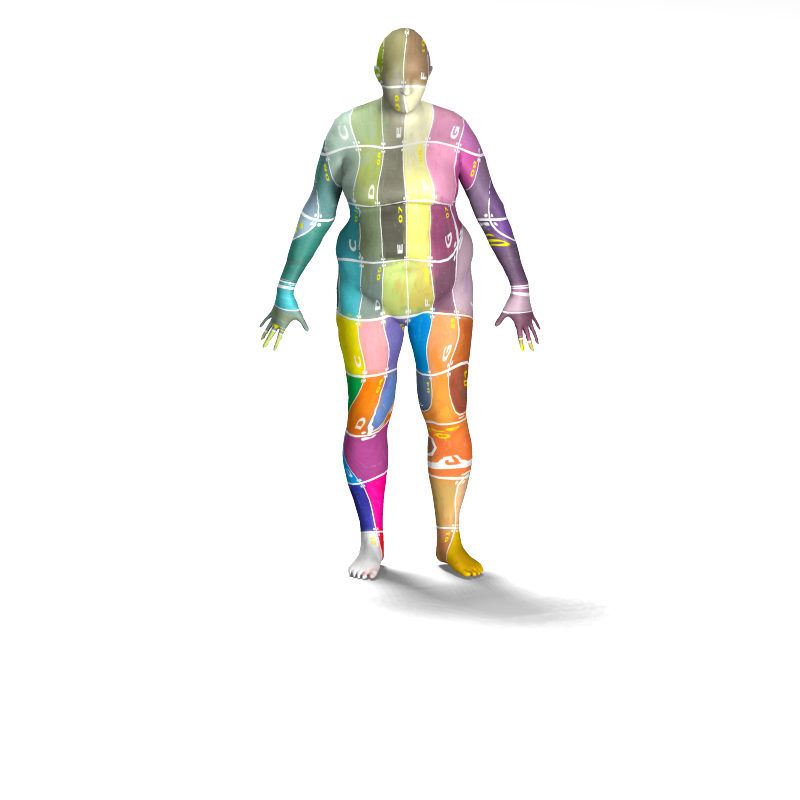}
  \caption*{FMNet + ICP}
  \label{fig:sfig2}
\end{subfigure}
\begin{subfigure}{.1\paperwidth}
  \centering
  \includegraphics[trim={9cm 4cm 8cm 0}, clip, width = \linewidth]{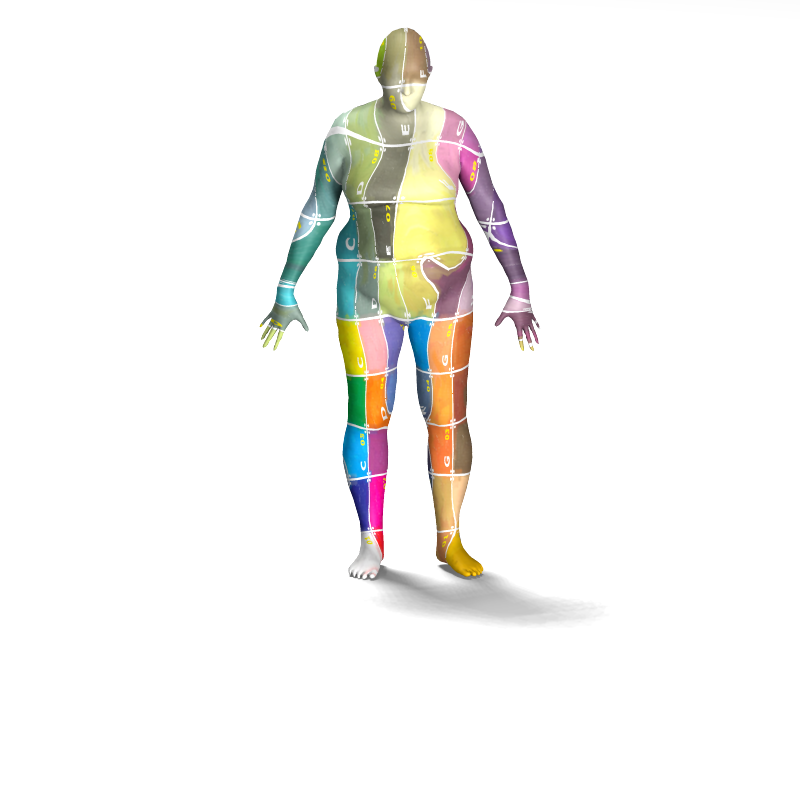}
  \caption*{GCNN}
  \label{fig:sfig2}
\end{subfigure}
\end{center}
\vspace{-4mm}
  \caption{Comparison of our method with \textit{Supervised} methods for texture transfer on the FAUST remeshed dataset.
  \vspace{-2mm}}
\label{fig:FAUST2}
\end{figure*}
\subsection{More Qualitative Comparison}\label{sup:morefigs} 
\begin{figure*}[t!]
\begin{center}
\begin{subfigure}{.1\paperwidth}
  \centering
  \includegraphics[trim={6cm 4cm 6cm 0}, clip, width=\linewidth]{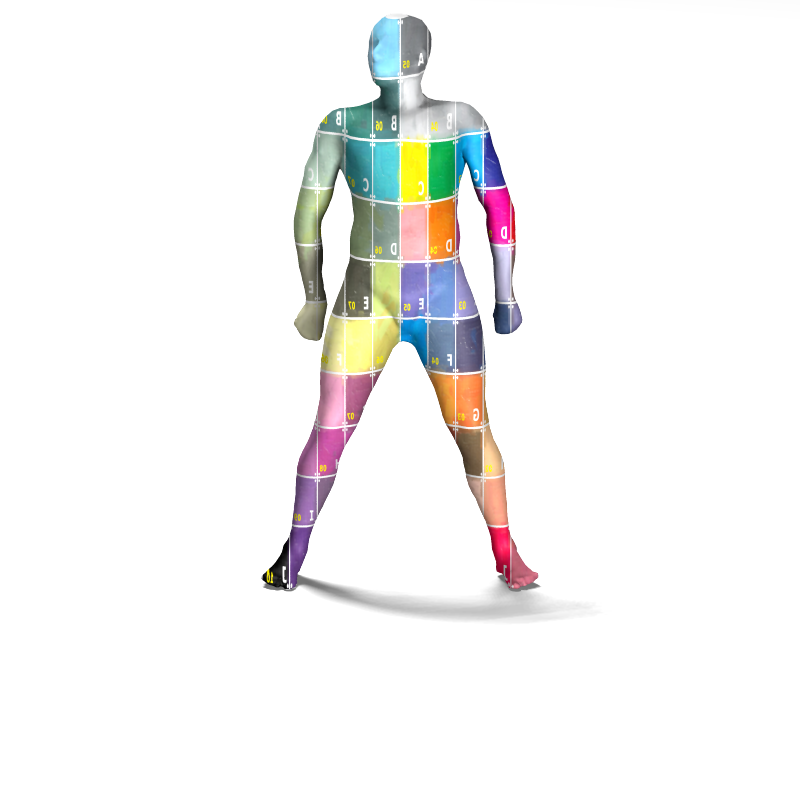}
  \caption*{Source}
  \label{fig:sfig1}
\end{subfigure}%
\begin{subfigure}{.1\paperwidth}
  \centering
  \includegraphics[trim={6cm 4cm 6cm 0}, clip, width = \linewidth]{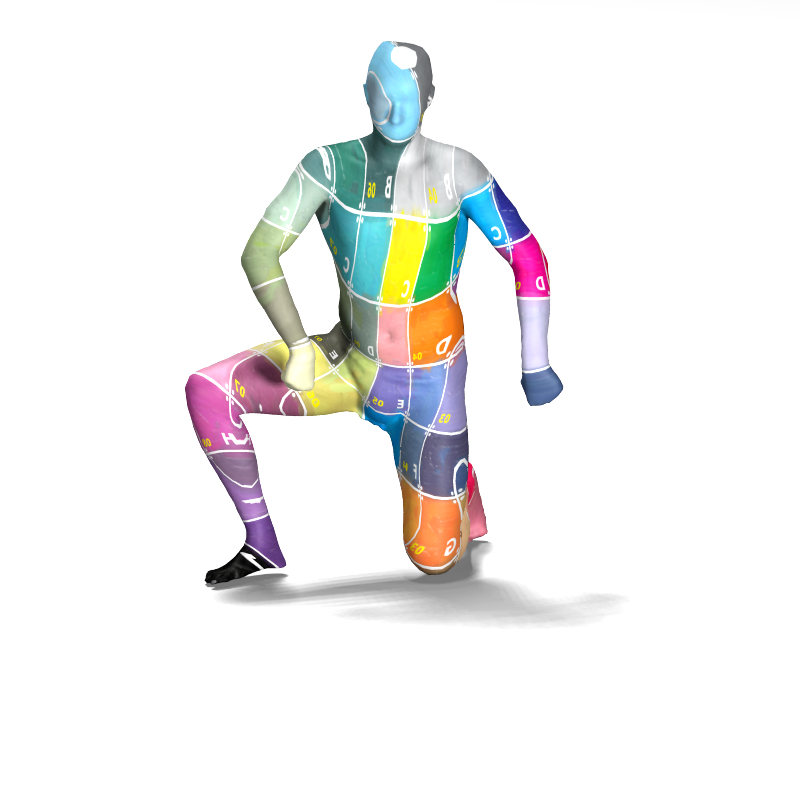}
  \caption*{Ground-Truth}
  \label{fig:sfig2}
\end{subfigure}
\begin{subfigure}{.12\paperwidth}
  \centering
  \includegraphics[trim={6cm 4cm 6cm 0}, clip, width = 0.9\linewidth]{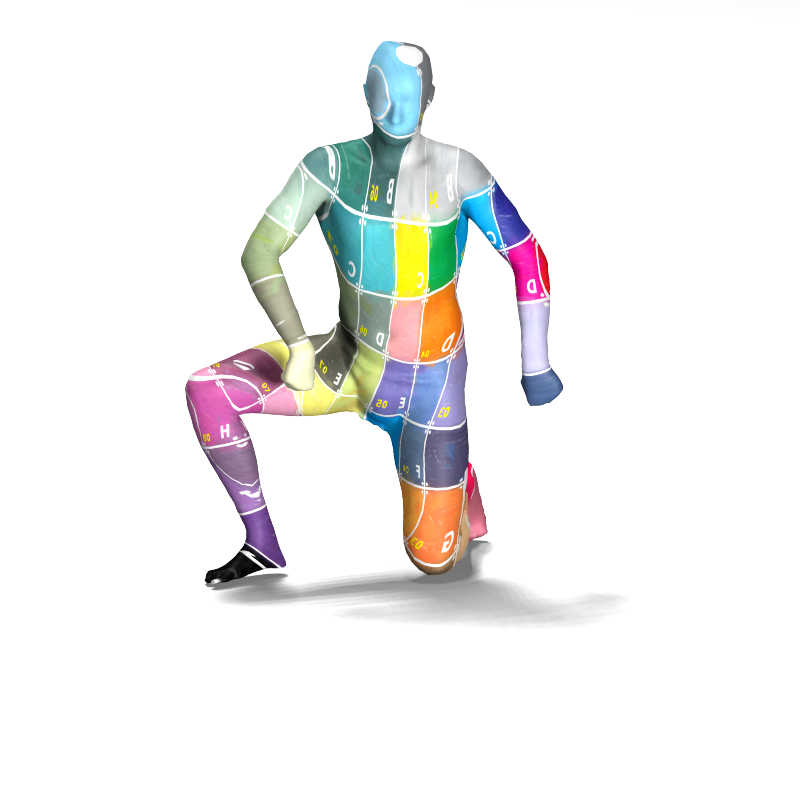}
  \vspace{-1mm}
  \caption*{SURFMNet + ICP}
  \label{fig:sfig2}
\end{subfigure}
\begin{subfigure}{.1\paperwidth}
  \centering
  \includegraphics[trim={6cm 4cm 6cm 0}, clip, width = \linewidth]{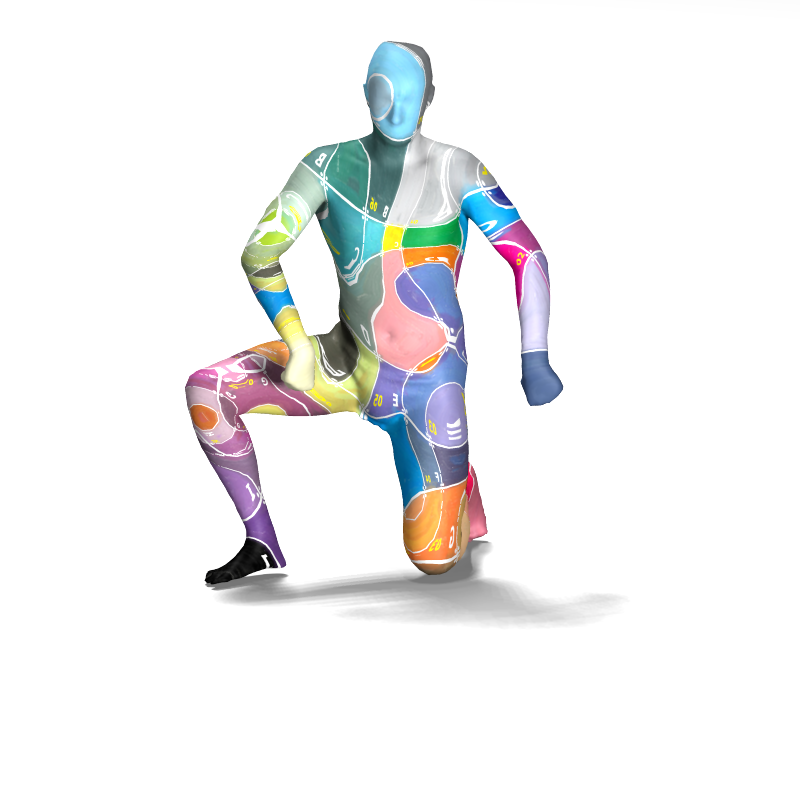}
  \caption*{SURFMNet}
  \label{fig:sfig2}
\end{subfigure}
\begin{subfigure}{.1\paperwidth}
  \centering
  \includegraphics[trim={6cm 4cm 6cm 0}, clip, width = \linewidth]{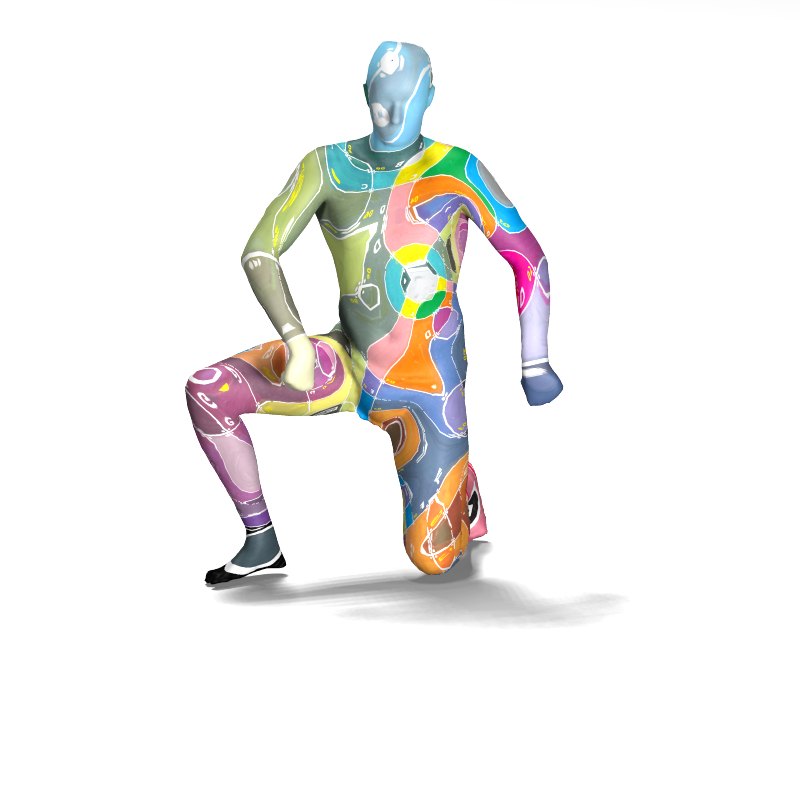}
  \caption*{FMNet}
  \label{fig:sfig2}
\end{subfigure}
\begin{subfigure}{.1\paperwidth}
  \centering
  \includegraphics[trim={6cm 4cm 6cm 0}, clip, width = \linewidth]{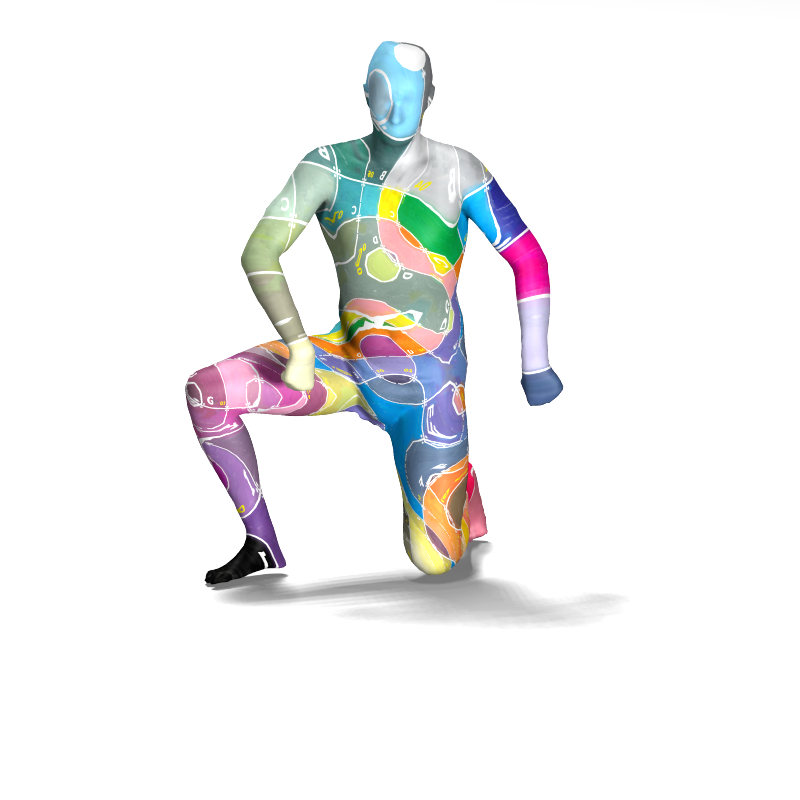}
  \caption*{FMNet + ICP}
  \label{fig:sfig2}
\end{subfigure}
\begin{subfigure}{.1\paperwidth}
  \centering
  \includegraphics[trim={6cm 4cm 6cm 0}, clip, width = \linewidth]{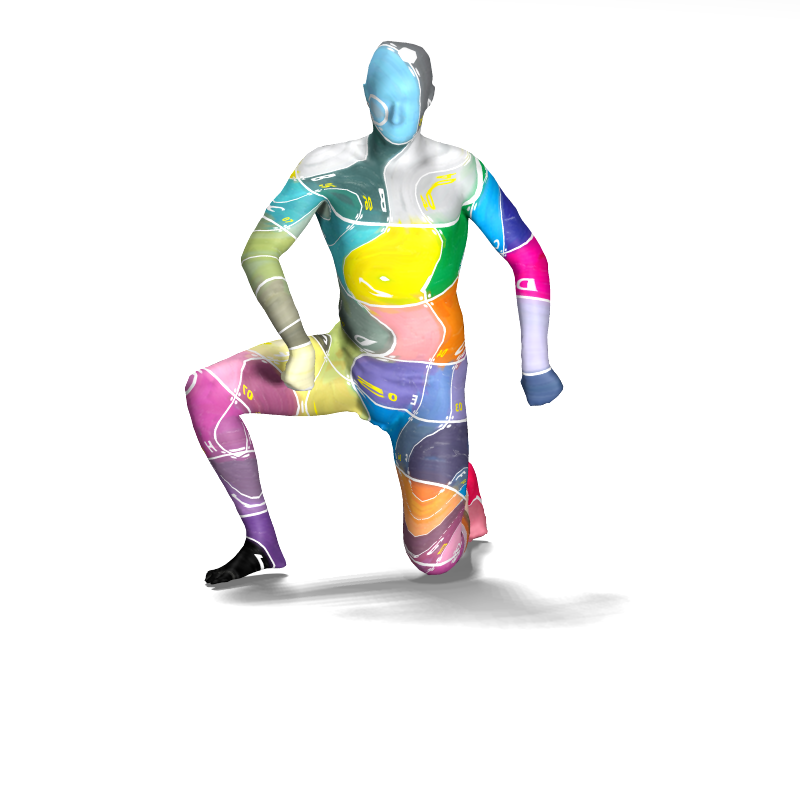}
  \caption*{GCNN}
  \label{fig:sfig2}
\end{subfigure}
\end{center}
\vspace{-4mm}
  \caption{Comparison of our method with \textit{Supervised} methods for texture transfer on the SCAPE remeshed dataset.
  \vspace{-2mm}}
\label{fig:SCAPE}
\end{figure*}

\begin{figure*}
\begin{center}
\begin{subfigure}{.128\paperwidth}
  \centering
  \includegraphics[trim={7cm 4cm 7cm 1cm},clip,width=\linewidth]{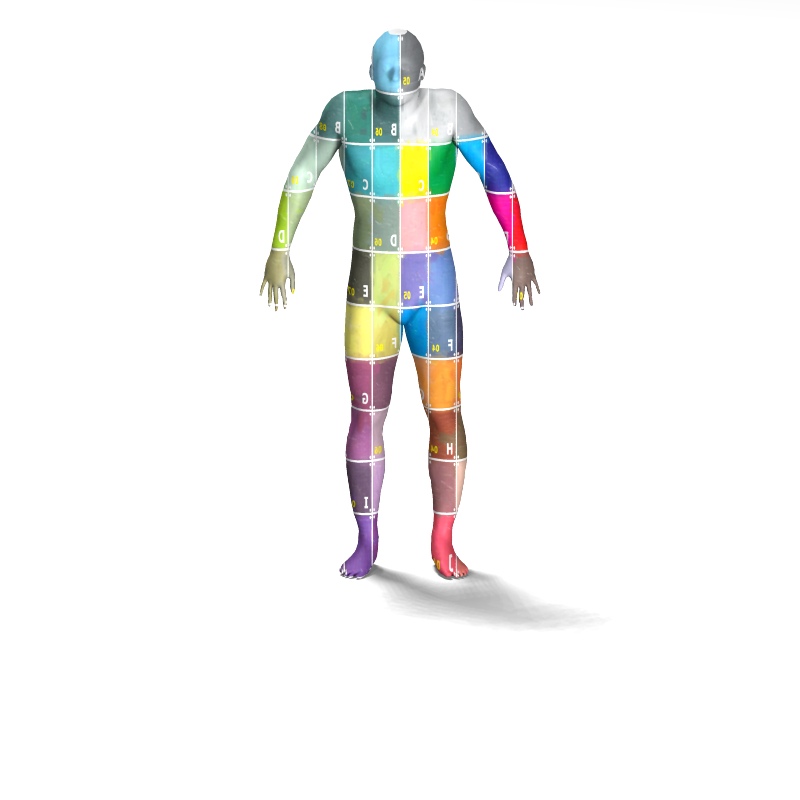}
  \caption*{Source}
  \label{fig:sfig1}
\end{subfigure}%
\begin{subfigure}{.128\paperwidth}
  \centering
  \includegraphics[trim={7cm 4cm 7cm 1cm},clip,width = \linewidth]{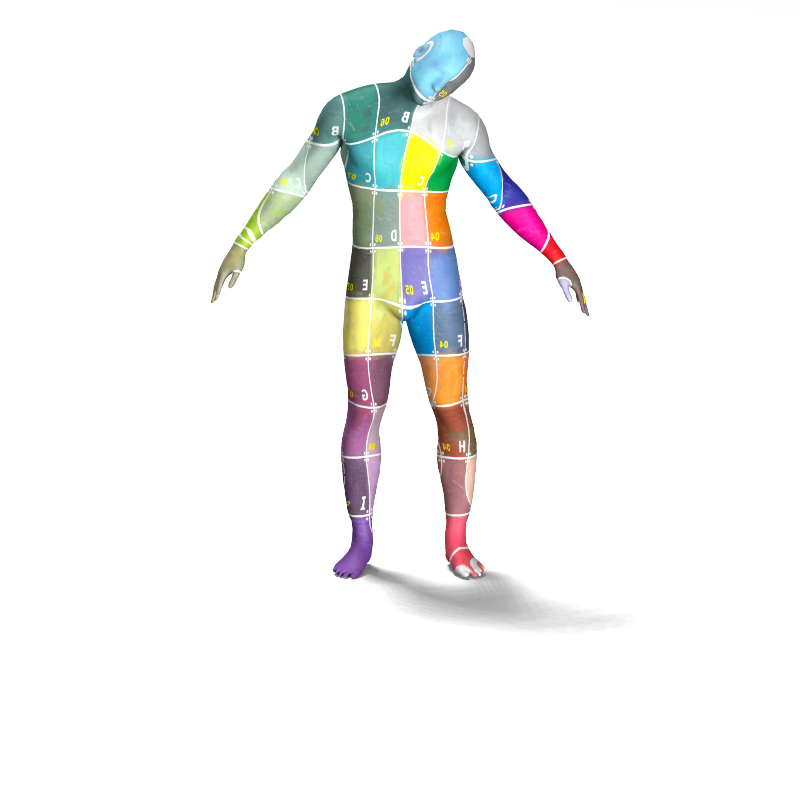}
  \caption*{Ground-Truth}
  \label{fig:sfig2}
\end{subfigure}
\begin{subfigure}{.128\paperwidth}
  \centering
  \includegraphics[trim={7cm 4cm 7cm 1cm}, clip, width = \linewidth]{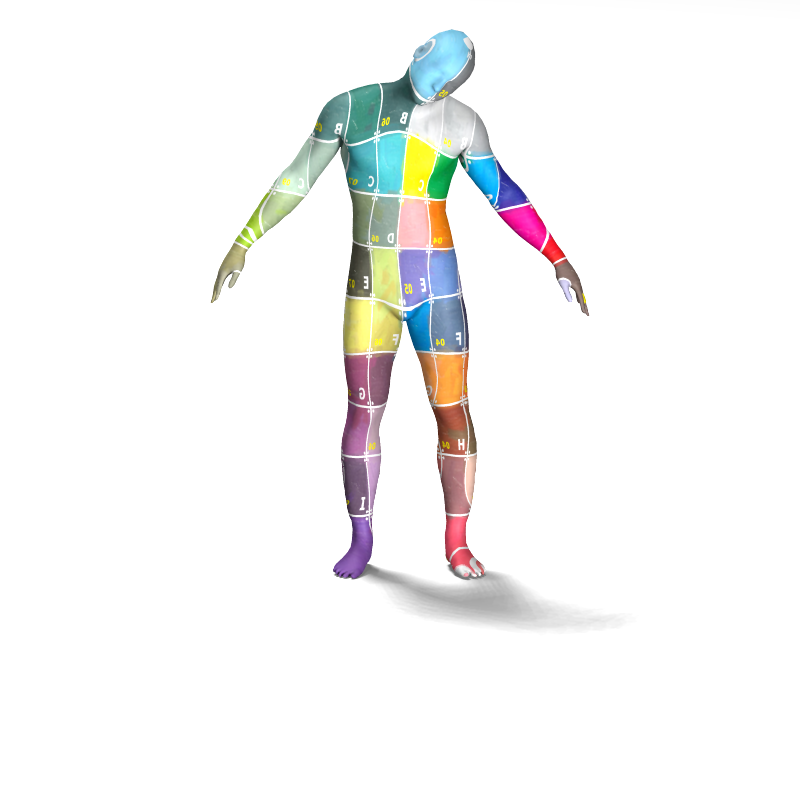}
  \caption*{SURFMNet}
  \label{fig:sfig2}
\end{subfigure}
\begin{subfigure}{.145\paperwidth}
  \centering
  \includegraphics[trim={0cm 0cm 0cm 0cm},clip,width = \linewidth]{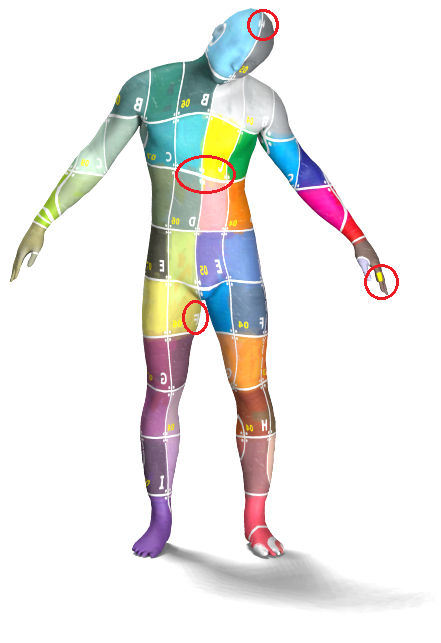}
  \caption*{BCICP}
  \label{fig:sfig2}
\end{subfigure}
\begin{subfigure}{.128\paperwidth}
  \centering
  \includegraphics[trim={7cm 4cm 7cm 1cm},clip,width = \linewidth]{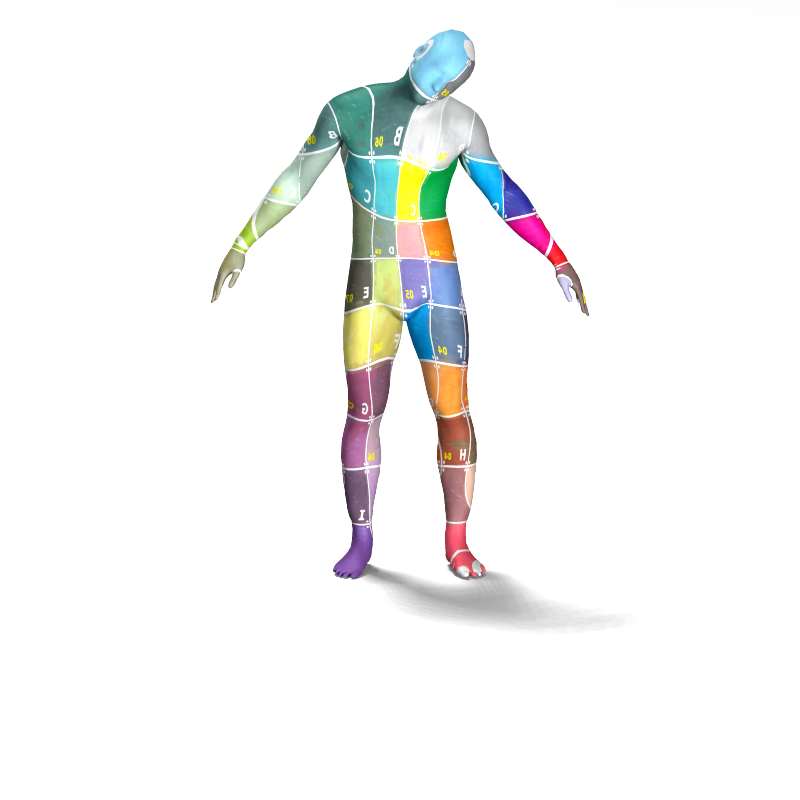}
  \caption*{PMF (heat)}
  \label{fig:sfig2}
\end{subfigure}
\begin{subfigure}{.128\paperwidth}
  \centering
  \includegraphics[trim={7cm 4cm 7cm 1cm},clip,width = \linewidth]{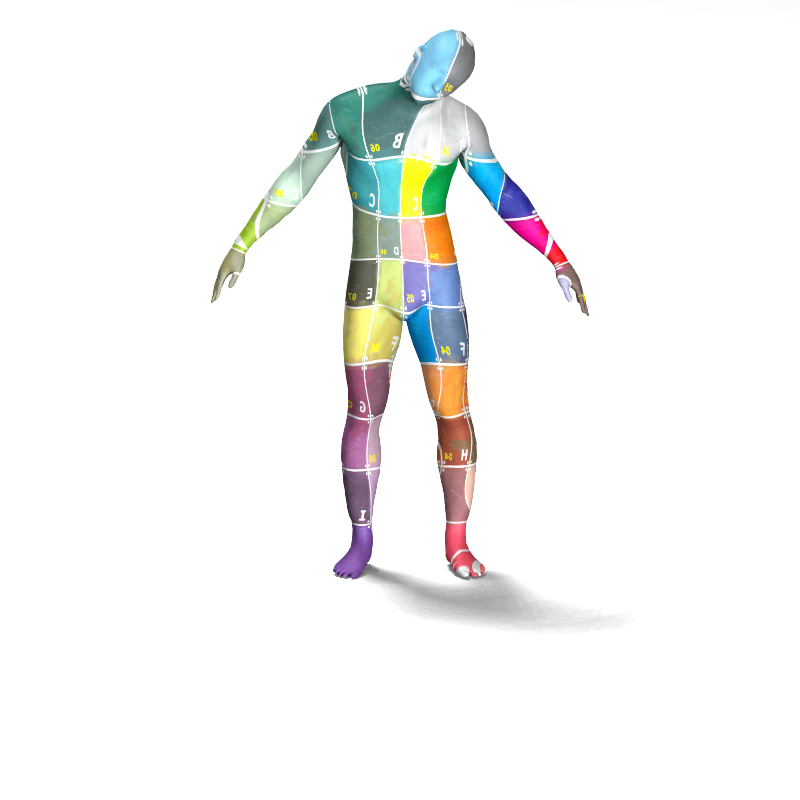}
  \caption*{PMF (gauss)}
  \label{fig:sfig2}
\end{subfigure}
\end{center}
\vspace{-4mm}
   \caption{Comparison of our method with \textit{Unsupervised} methods for texture transfer on the FAUST remeshed dataset. Note that BCICP is roughly 7 times slower when compared to our method. We highlight the shortcomings of BCICP matching with red circles.
   \vspace{-2mm}}
\label{fig:FAUST}
\end{figure*}

In Figures \ref{fig:FAUST2} and  \ref{fig:FAUST} , we provide more qualitative comparisons of SURFMNet on the FAUST remeshed datasets whereas Figures $10$ and \ref{fig:SCAPE} provide a comparison on the SCAPE remeshed dataset. In all cases, our method produces the highest quality maps.

\begin{figure*}[t!]
\begin{center}
\begin{subfigure}{.13\paperwidth}
  \centering
  \includegraphics[trim={8cm 5cm 6cm 0}, clip,  width=\linewidth]{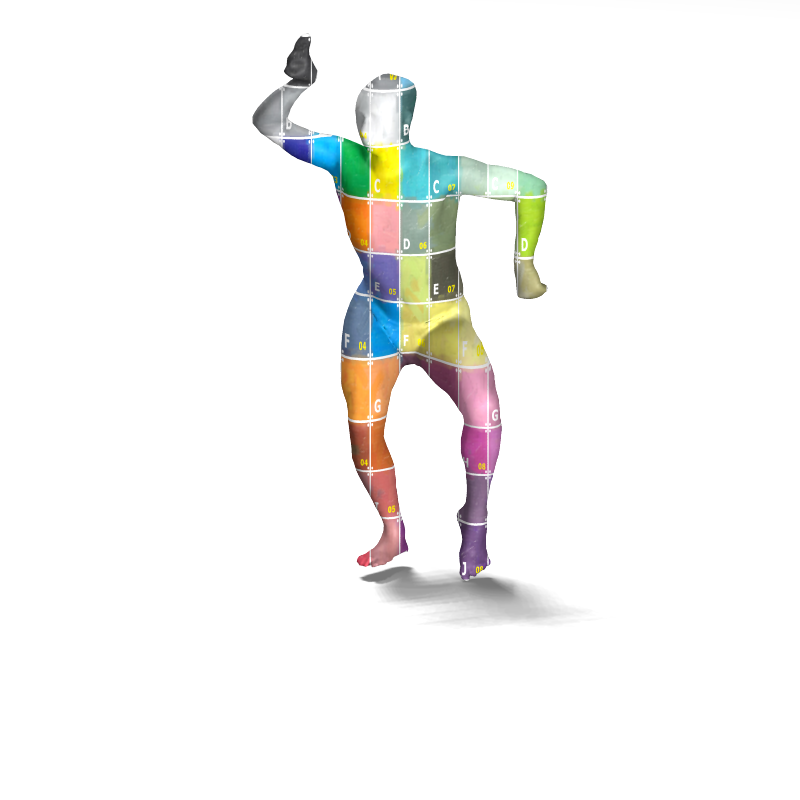}
  \caption*{Source}
  \label{fig:sfig1}
\end{subfigure}%
\begin{subfigure}{.13\paperwidth}
  \centering
  \includegraphics[trim={8cm 5cm 6cm 0}, clip, width = \linewidth]{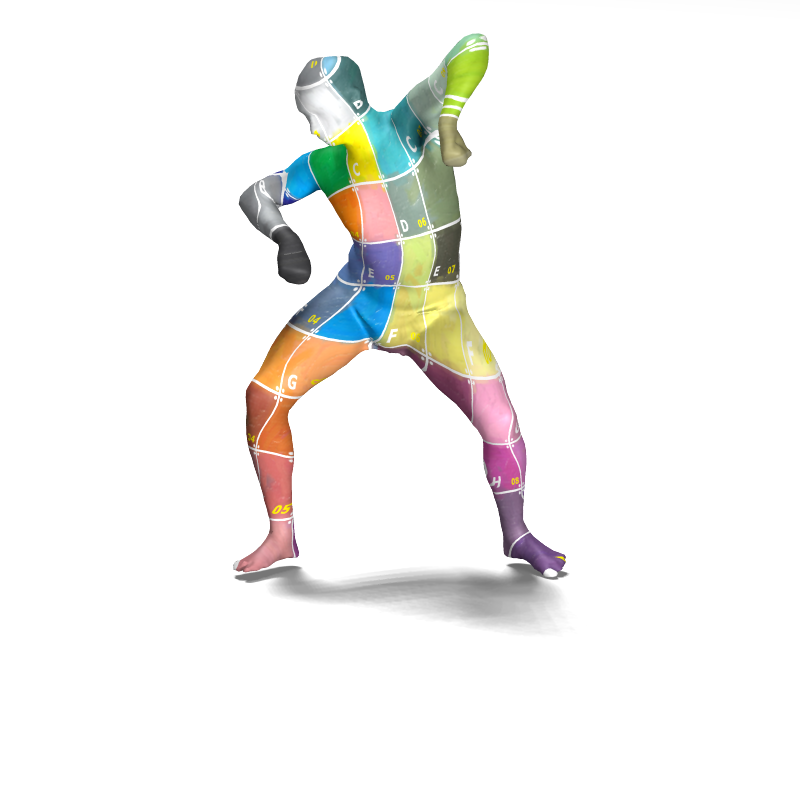}
  \caption*{Ground-Truth}
  \label{fig:sfig2}
\end{subfigure}
\begin{subfigure}{.13\paperwidth}
  \centering
  \includegraphics[trim={8cm 5cm 6cm 0}, clip, width = \linewidth]{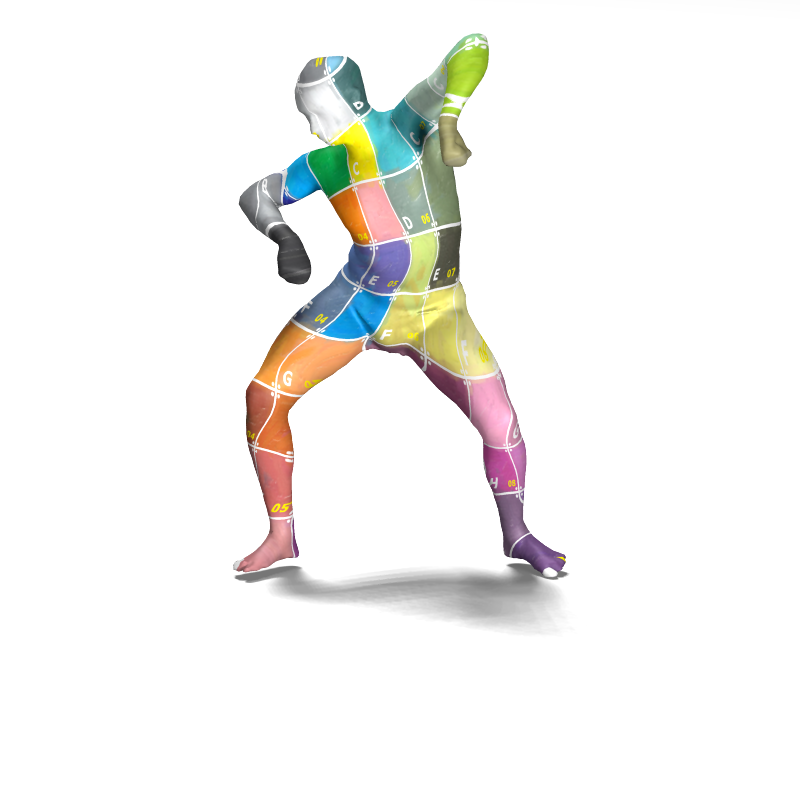}
  \caption*{SURFMNet-all}
  \label{fig:sfig2}
\end{subfigure}
\begin{subfigure}{.13\paperwidth}
  \centering
  \includegraphics[trim={8cm 5cm 6cm 0}, clip, width = \linewidth]{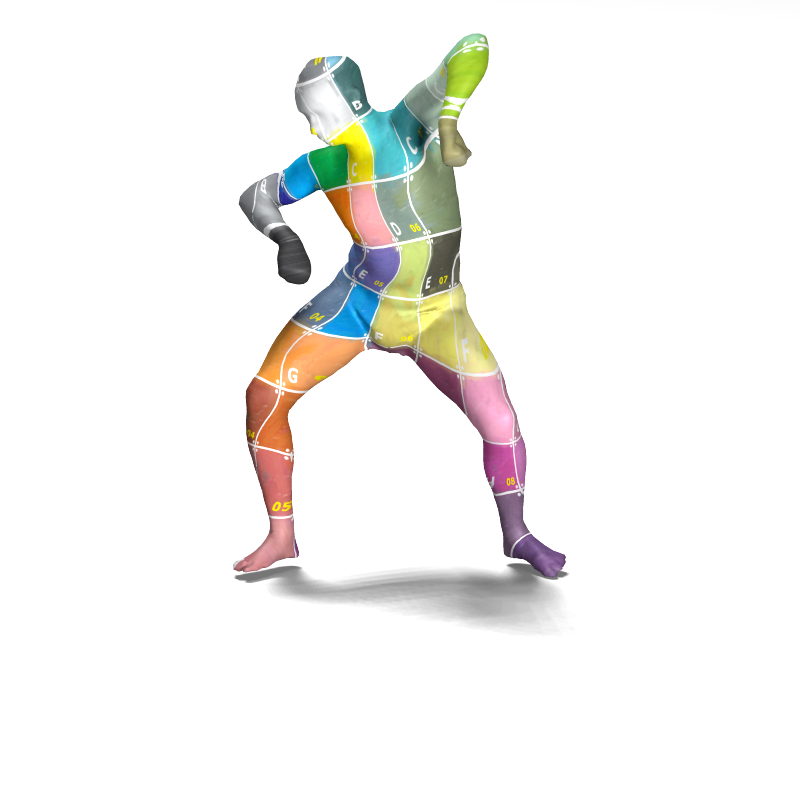}
  \caption*{BCICP}
  \label{fig:sfig2}
\end{subfigure}
\begin{subfigure}{.13\paperwidth}
  \centering
  \includegraphics[trim={8cm 5cm 6cm 0}, clip, width = \linewidth]{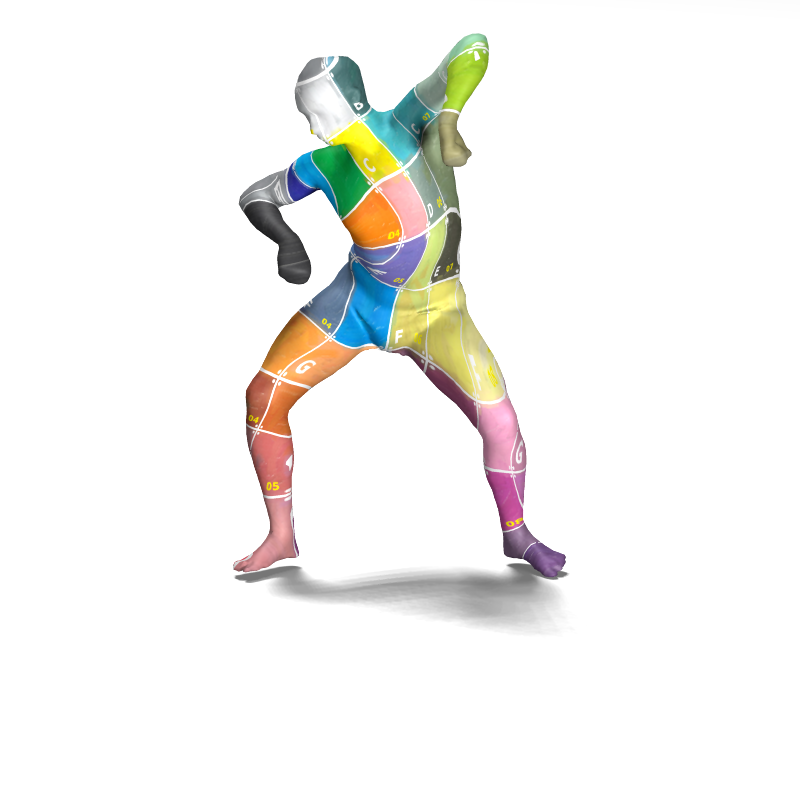}
  \caption*{PMF (heat)}
  \label{fig:sfig2}
\end{subfigure}
\begin{subfigure}{.13\paperwidth}
  \centering
  \includegraphics[trim={8cm 5cm 6cm 0}, clip, width = \linewidth]{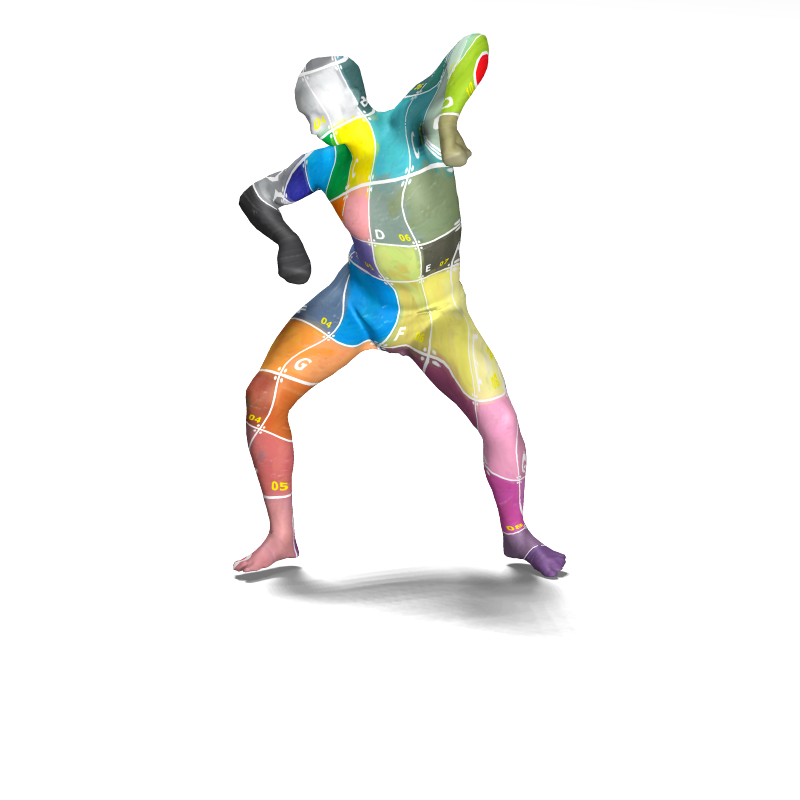}
  \caption*{PMF (gauss)}
  \label{fig:sfig2}
\end{subfigure}
\end{center}
\vspace{-5mm}
   \caption{Comparison of our method with \textit{Unsupervised} methods for texture transfer on the SCAPE remeshed dataset.
   \vspace{-2mm}}
\label{fig:unsup}
\end{figure*}

\end{document}